\documentclass[prd,superscriptaddress,unsortedaddress,twocolumn,showpacs,floatfix]{revtex4}

\usepackage{epsfig}
\usepackage{dcolumn}
\usepackage{amsmath}

\newcommand{\rklrad}{\mbox{${\cal R}_{K\ell 3\gamma}$}}
\newcommand{\rkerad}{\mbox{${\cal R}_{K e 3\gamma}$}}
\newcommand{\rkmurad}{\mbox{${\cal R}_{K \mu 3\gamma}$}}

\newcommand{\epsrat}{\mbox{$\epsilon^{\prime}\!/\epsilon$}}
\newcommand{\reepoe}{\mbox{$Re(\epsrat)$}}

\newcommand{\KLpilnu}{\mbox{$K_{L}\to\pi^{\pm}\ell^{\mp}\nu$}}
\newcommand{\Klthree}{\mbox{$K_{\ell 3}$}}
\newcommand{\Klthreerad}{\mbox{$K_{\ell 3\gamma}$}}
\newcommand{\Kethreerad}{\mbox{$K_{e 3\gamma}$}}
\newcommand{\Kmuthreerad}{\mbox{$K_{\mu 3\gamma}$}}
\newcommand{\KLSpp}{\mbox{$K_{L,S}\to\pi^+\pi^-$}}
\newcommand{\ppzkin}{\mbox{$k_{+-0}$}}
\newcommand{\Kthreepi}{\mbox{$K_{3\pi}$}}
\newcommand{\Kethree}{\mbox{$K_{e3}$}}
\newcommand{\Kmuthree}{\mbox{$K_{\mu 3}$}}
\newcommand{\KLpienu}{\mbox{$K_{L}\to\pi^{\pm}e^{\mp}\nu$}}
\newcommand{\KLpimunu}{\mbox{$K_{L}\to\pi^{\pm}{\mu}^{\mp}\nu$}}

\newcommand{\KLzzz}{\mbox{$K_{L}\rightarrow \pi^0\pi^0\pi^0$}}
\newcommand{\KLzz}{\mbox{$K_{L}\rightarrow \pi^0\pi^0$}}
\newcommand{\KLpmz}{\mbox{$K_{L}\rightarrow \pi^+\pi^-\pi^0$}}
\newcommand{\ppc}{\mbox{$\pi^+\pi^-$}}
\newcommand{\pz}{\mbox{$\pi^0$}}

\newcommand{\KLpipienu}{\mbox{$K_{L}\to\pi^0\pi^{\pm}e^{\mp}\nu$}}
\newcommand{\KLpipimunu}{\mbox{$K_{L}\to\pi^0\pi^{\pm}\mu^{\mp}\nu$}}

\newcommand{\gklthree}{\mbox{$\Gamma_{K\ell 3}$ }}

\newcommand{\gklborn}{\mbox{$\Gamma_{K\ell 3}^{\rm Born}$ }}

\newcommand{\ktev}{\mbox{KTeV}}
\newcommand{\Egcm}{\mbox{$E_{\gamma}^{\star}$}}
\newcommand{\thgamlepcm}{\mbox{$\theta_{\gamma\ell}^{\star}$}}
\newcommand{\thgamecm}{\mbox{$\theta_{\gamma e}^{\star}$}}
\newcommand{\thgammucm}{\mbox{$\theta_{\gamma \mu}^{\star}$}}
\newcommand{\pihad}{$\pi$-hadron interaction}
\newcommand{\brems}{bremsstrahlung}
\newcommand{\drbrem}{\mbox{${\Delta R}_{\gamma{\rm -brem} }$}}
\newcommand{\upk}{ {\rm GeV}/c }
\newcommand{\schi}{ \mbox{{\rm shape-}$\chi_{\gamma}^2$ }}
\newcommand{\magvus}{\mbox{$|V_{us}|$}}

\newcommand{\ETWO}{\mbox{$\eta_{\gamma\gamma}$}}
\newcommand{\RADRAD}{\mbox{$A_{\gamma\gamma/\gamma}$}}
\newcommand{\RADRADVALUE}{0.982}
\newcommand{\RADRADERRABS}{0.005}
\newcommand{\RADRADERRPERCENT}{0.5}

\newcommand{\radcor}{\mbox{$\delta^\ell_K$}}
\newcommand{\RADCOR}{\mbox{$\delta_{rad}$}}

\newcommand{\NDATA}{ N_{mode}^{data} }
\newcommand{\NMCREC}{ N_{mode}^{MCrec} }
\newcommand{\NMCRECKL}{ N_{K\ell 3}^{MCrec} }
\newcommand{\NMCRECKLg}{ N_{K\ell 3\gamma}^{MCrec} }

\newcommand{\NMCGEN}{ N_{mode}^{MCgen} }
\newcommand{\NMCGENKL}{ N_{K\ell 3}^{MCgen} }
\newcommand{\NMCGENKLg}{ N_{K\ell 3\gamma}^{MCgen} }

\def\KEbgke4{ ( 0.07 \pm  0.03) }
\def\KEbgpihad{ ( 0.54 \pm  0.11) }
\def\KEbgaccid{ ( 0.22 \pm  0.11) }
\def\KEbgxbrem{ ( 0.20 \pm  0.02) }
\def\KEbgsum{ ( 1.04 \pm  0.16) }
\def\KEBGsyst{  0.16}
 
\def\KMUbgkethree{ ( 0.06 \pm  0.02) }
\def\KMUbgkpmz{ ( 4.39 \pm  1.75) }
\def\KMUbgkpmg{ ( 0.19 \pm  0.08) }
\def\KMUbgkmu4{  <  0.05}
\def\KMUbgpihad{ ( 1.60 \pm  0.32) }
\def\KMUbgaccid{ ( 1.68 \pm  0.89) }
\def\KMUbgsum{ ( 7.91 \pm  1.99) }
\def\KMUBGsyst{  1.99}
 
\def\KMUXCLVTO{  0.60}
\def\KEXCLVTO{  0.58}
 
\def\NKethree{ 2191077}
\def\NKethreerad{   14221}
\def\NKmuthree{ 1691400}
\def\NKmuthreerad{    1385}
 
\def\RKERAD{   4.942}
\def\RKERADerrstat{   0.042}
\def\RKERADerrsyst{   0.046}
\def\RKERADerrtot{   0.062}
\def\RKERADrelstat{  0.84}
\def\RKERADrelsyst{  0.92}
\def\RKERADreltot{  1.25}
\def\RKERADmcstat{  0.28}
 
\def\RKMURAD{   0.530}
\def\RKMURADerrstat{   0.014}
\def\RKMURADerrsyst{   0.012}
\def\RKMURADerrtot{   0.019}
\def\RKMURADrelstat{  2.69}
\def\RKMURADrelsyst{  2.36}
\def\RKMURADreltot{  3.58}
\def\RKMURADmcstat{  0.89}
 
\def\OLDKMURAD{   0.209}

\def\OLDKMURADerrtot{   0.009}
 
\def\OLDKERAD{   0.916}

\def\OLDKERADerrtot{   0.017}
 
\def\FAKERADDATA{ ( 7.62 \pm  0.40) }
\def\FAKERADMC{ ( 8.06 \pm  0.29) }

\def\KLORKERAD{ 4.937 \pm 0.067 }
\def\KLOROLDKERAD{ 0.956 \pm 0.013 }
\def\KLORKMRAD{ 0.564 \pm 0.011 }
\def\KLOROLDKMRAD{ 0.214 \pm 0.004 }
 
\def\KERADRATIO{ 1.001 \pm 0.018 }
\def\OLDKERADRATIO{ 0.958 \pm 0.022 }
\def\KMRADRATIO{ 0.938 \pm 0.038 }
\def\OLDKMRADRATIO{ 0.979 \pm 0.046 }

\begin{document}     


\title{ 
       Measurements of the Branching Fraction and \\
       Decay Distributions
       for $\KLpimunu\gamma$ and $\KLpienu\gamma$.
           }

\newcommand{\UAz}{University of Arizona, Tucson, Arizona 85721}
\newcommand{\UCLA}{University of California at Los Angeles, Los Angeles,
                    California 90095} 
\newcommand{\UCSD}{University of California at San Diego, La Jolla,
                   California 92093} 
\newcommand{\EFI}{The Enrico Fermi Institute, The University of Chicago, 
                  Chicago, Illinois 60637}
\newcommand{\UB}{University of Colorado, Boulder, Colorado 80309}
\newcommand{\ELM}{Elmhurst College, Elmhurst, Illinois 60126}
\newcommand{\FNAL}{Fermi National Accelerator Laboratory, 
                   Batavia, Illinois 60510}
\newcommand{\Osaka}{Osaka University, Toyonaka, Osaka 560-0043 Japan} 
\newcommand{\Rice}{Rice University, Houston, Texas 77005}
\newcommand{\UVa}{The Department of Physics and Institute of Nuclear and 
                  Particle Physics, University of Virginia, 
                  Charlottesville, Virginia 22901}
\newcommand{\UW}{University of Wisconsin, Madison, Wisconsin 53706}

\affiliation{\UAz}
\affiliation{\UCLA}
\affiliation{\UCSD}
\affiliation{\EFI}
\affiliation{\UB}
\affiliation{\ELM}
\affiliation{\FNAL}
\affiliation{\Osaka}
\affiliation{\Rice}
\affiliation{\UVa}
\affiliation{\UW}

\author{T.~Alexopoulos}   \affiliation{\UW}
\author{T.~Andre}         \affiliation{\EFI}
\author{M.~Arenton}       \affiliation{\UVa}
\author{R.F.~Barbosa}     \altaffiliation[Permanent address: ]
   {University of S\~{a}o Paulo, S\~{a}o Paulo, Brazil}\affiliation{\FNAL}
\author{A.R.~Barker}      \altaffiliation[Deceased.]{ } \affiliation{\UB}
\author{L.~Bellantoni}    \affiliation{\FNAL}
\author{A.~Bellavance}    \affiliation{\Rice}
\author{E.~Blucher}       \affiliation{\EFI}
\author{G.J.~Bock}        \affiliation{\FNAL}
\author{E.~Cheu}          \affiliation{\UAz}
\author{S.~Childress}     \affiliation{\FNAL}
\author{R.~Coleman}       \affiliation{\FNAL}
\author{M.D.~Corcoran}    \affiliation{\Rice}
\author{B.~Cox}           \affiliation{\UVa}
\author{A.R.~Erwin}       \affiliation{\UW}
\author{R.~Ford}          \affiliation{\FNAL}
\author{A.~Glazov}        \affiliation{\EFI}
\author{A.~Golossanov}    \affiliation{\UVa}
\author{J.~Graham}        \affiliation{\EFI}   
\author{J.~Hamm}          \affiliation{\UAz}
 
\author{K.~Hanagaki}      \affiliation{\Osaka}
\author{Y.B.~Hsiung}      \affiliation{\FNAL}
\author{H.~Huang}         \affiliation{\UB}
\author{V.~Jejer}         \affiliation{\UVa}  
\author{D.A.~Jensen}      \affiliation{\FNAL}
\author{R.~Kessler}       \affiliation{\EFI}
\author{H.G.E.~Kobrak}    \affiliation{\UCSD}
\author{K.~Kotera}        \affiliation{\Osaka}
\author{J.~LaDue}         \affiliation{\UB}
  
\author{A.~Ledovskoy}     \affiliation{\UVa}
\author{P.L.~McBride}     \affiliation{\FNAL}

\author{E.~Monnier}
   \altaffiliation[Permanent address: ]{C.P.P. 
    Marseille/C.N.R.S., France}\affiliation{\EFI}
\author{K.S.~Nelson}       \affiliation{\UVa}
\author{H.~Nguyen}       \affiliation{\FNAL}
\author{R.~Niclasen}     \affiliation{\UB} 
\author{V.~Prasad}       \affiliation{\EFI}
\author{X.R.~Qi}         \affiliation{\FNAL}
\author{E.J.~Ramberg}    \affiliation{\FNAL}
\author{R.E.~Ray}        \affiliation{\FNAL}
\author{M.~Ronquest}	 \affiliation{\UVa}
\author{E. Santos}       \altaffiliation[Permanent address: ]
      {University of S\~{a}o Paulo, S\~{a}o Paulo, Brazil}\affiliation{\FNAL}
\author{P.~Shanahan}     \affiliation{\FNAL}
\author{J.~Shields}      \affiliation{\UVa}
\author{W.~Slater}       \affiliation{\UCLA}
\author{D.~Smith}	 \affiliation{\UVa}
\author{N.~Solomey}      \affiliation{\EFI}
\author{E.C.~Swallow}    \affiliation{\EFI}\affiliation{\ELM}
\author{P.A.~Toale}      \affiliation{\UB}
\author{R.~Tschirhart}   \affiliation{\FNAL}
\author{Y.W.~Wah}        \affiliation{\EFI}
\author{J.~Wang}         \affiliation{\UAz}
\author{H.B.~White}      \affiliation{\FNAL}
\author{J.~Whitmore}     \affiliation{\FNAL}
\author{M.~Wilking}      \affiliation{\UB}
\author{B.~Winstein}     \affiliation{\EFI}
\author{R.~Winston}      \affiliation{\EFI}
\author{E.T.~Worcester}  \affiliation{\EFI}
\author{T.~Yamanaka}     \affiliation{\Osaka}
\author{E.~D.~Zimmerman} \affiliation{\UB}

\collaboration{The KTeV Collaboration}


\date{\today}

\begin{abstract}
  We present measurements of
  $\rklrad \equiv
       \Gamma(\KLpilnu\gamma;~\Egcm >10~{\rm MeV})/\Gamma(\KLpilnu)$,
  where $\ell = \mu$ or $e$, and $\Egcm$ is the photon energy in the
  kaon rest frame.
  These measurements are based on $K_L$ decays collected in 1997
  by the \ktev\ (E832) experiment at Fermilab.
  With samples of 
    $\NKmuthreerad$ $\KLpimunu\gamma$ and 
    $\NKethreerad$  $\KLpienu\gamma$ candidates,
  we find 
  $\rkmurad = (\RKMURAD \pm \RKMURADerrtot)$\% and
  $\rkerad  = (\RKERAD  \pm \RKERADerrtot)$\%.
  We also examine distributions
  of photon energy and lepton-photon angle.
\end{abstract}

\pacs{13.25.Es, 14.40.Aq}

\maketitle


%
%

 \section{Introduction}
 \label{sec:intro}

Radiative effects play an important role in relating the 
Cabibbo-Kobayashi-Maskawa (CKM) parameter
\magvus\ to $K_L$ semileptonic decays.
Radiative effects enter the measurement of \magvus\
in two distinct ways.
First, \magvus\ is extracted from the semileptonic decay rate
$\gklthree = \gklborn \times (1+\RADCOR)$,
where $\ell=e$ or $\mu$,
$\gklborn$ is the Born-level decay rate proportional
to $\magvus^2$,
and $\RADCOR$ describes radiative corrections.
In our extraction of $\magvus$~\cite{prl_vus}
$(1+\RADCOR)= S_{EW}(1+\radcor)$,
where $S_{EW}$ is the short-distance radiative correction 
taken from~\cite{sew}, and the {\sc klor} program~\cite{Troy}
is used to determine $\radcor$.
The second role of radiative effects is in the 
Monte Carlo simulation (MC) that is used to determine 
the detector acceptance needed
to measure the \Klthree\ branching fractions and form factors.
A Monte Carlo simulation requires a precise understanding of the
kinematic distributions, particularly the photon energy spectrum
and the angle between the charged lepton and photon.
The sensitivity of the MC to radiative effects depends
on the experimental technique.
For the \ktev\ measurements \cite{prd_br,prd_ff},
radiative effects in the simulation change the \Kethree\ 
branching fraction by a few percent, 
and have a more significant impact on the form factors.
The {\sc klor} program that is used to determine $\radcor$
is also used to generate $\KLpilnu(\gamma)$ decays in our 
Monte Carlo simulation, and to predict radiative semileptonic
branching fractions.

To test our understanding of radiative semileptonic decays,
we measure
\begin{equation}
   \rklrad \equiv  
      \frac{ \Gamma(\KLpilnu\gamma;~\Egcm > 10~{\rm MeV}) }
           { \Gamma(\KLpilnu) } ~,
   \label{eq:rkpilnurad}
\end{equation}
where $\ell = e$ or $\mu$, and $\Egcm$ is the photon energy in
the kaon rest frame. 
The denominator in Eq.~(\ref{eq:rkpilnurad}) is the \Klthree\
decay rate which includes the emission of one or more radiated photons
\footnote{
  \Klthree\ refers to semileptonic decays including radiation;
  \Klthreerad\ refers to radiative $\KLpilnu\gamma$ in which the
  photon energy is above the analysis threshold of 10~MeV or 30~MeV.
    }.
We also measure distributions of $\Egcm$ and 
the angle ($\thgamlepcm$) between the photon 
and charged lepton.

Previous \Klthreerad\ analyses selected $\Egcm > 30$~MeV;
for \Kethreerad, an additional angular cut, $\thgamecm > 20^0$,
was also required.
In this analysis, we select events with  $\Egcm > 10$~MeV
and remove the $\thgamlepcm$ requirement;
these relaxed cuts allow us to study 5~times more radiated
photons for $\KLpienu\gamma$,
and 2.5~times more photons for $\KLpimunu\gamma$.

The outline of this paper is as follows.
Section~\ref{sec:phenom} describes the theoretical
treatment used to generate radiative 
effects in semileptonic decays.
The \ktev\ apparatus is described in Sec.~\ref{sec:apparatus},
the \rklrad\  analyses are presented in Sec.~\ref{sec:anal},
and systematic uncertainties are discussed in Sec.~\ref{sec:syst}.
Section~\ref{sec:extraction} describes the acceptance
correction, as well as a correction from second-order
radiative effects.
Finally, results are presented in Sec.~\ref{sec:results}.

 \section{Treatment of Radiative Decays}
 \label{sec:phenom}

In our Monte Carlo simulation, 
we use the {\sc klor} program to generate both radiative
and non-radiative \Klthree\ events.  
{\sc klor} models first-order radiative corrections to the 
\Klthree\ decay mode using a phenomenological model
\cite{Troy,Ginsberg,FFS};
below we give a brief description of this program.
Second-order radiative effects are estimated using the
{\sc photos} program~\cite{photos2}, and are
discussed in Sec.~\ref{subsec:rad2}.

\begin{figure}[hb]
\centering
\epsfig{file=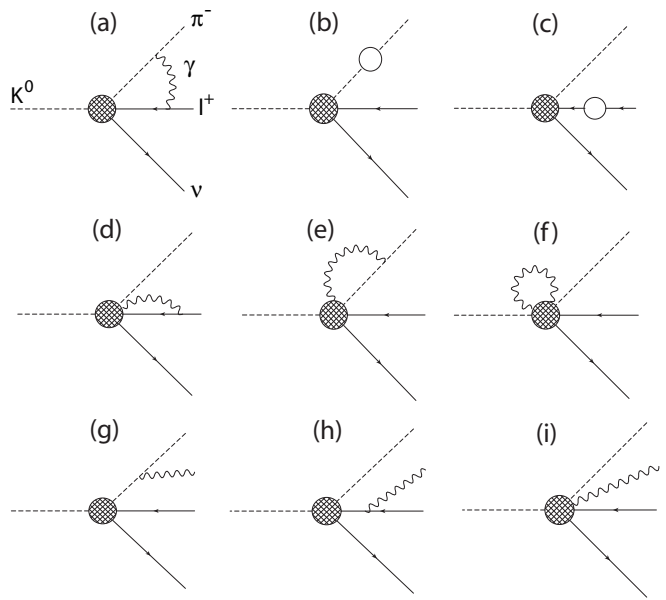,width=\linewidth}
\caption{\label{fig:feynman} 
     Feynman diagrams for the first-order
     radiative corrections to the $K_{\ell 3}$ decay mode.  
     Diagrams (a)--(f) are the virtual corrections while 
     diagrams (g)--(i) are the inner-\brems\ corrections.  
     The open circles in diagram (b) and (c) denote the self-energy 
     correction to the pion and lepton wavefunctions, respectively.
                }
\end{figure}

First-order radiative corrections are composed of both 
inner-\brems\ (IB) and virtual contributions.
Figure~\ref{fig:feynman}(g)--(i) shows the IB 
contribution consisting of radiation from the pion,
the charged lepton, and the 
vertex~\footnote{Radiation from the
vertex is required to preserve gauge invariance.}.
Data-MC comparisons of the kinematic distributions for 
radiated photons ($\Egcm$ and $\thgamlepcm$) 
are presented in Sec.~\ref{subsec:discuss}.
Figure~\ref{fig:feynman}(a)--(f) illustrates the 
virtual corrections that involve the emission and absorption of
a photon by the pion, the charged lepton, or the effective vertex.
The effects of virtual photon exchange are most prominent in the 
distributions of the pion-lepton mass ($m_{\pi\ell}$), 
and the transverse momentum of each charged particle ($p_T$).  
Using {\sc klor}, data-MC comparisons in these distributions 
show good agreement (Fig.~3 and Fig.~4 in ~\cite{prd_ff}).

Radiative corrections to \Klthree\
decays depend on the hadronic $K$--$\pi$ form factors.
The form factors depend on $t$, the square of the
four-momentum transfer to the gauge boson ($W^{\pm}$).
This four-momentum squared is given by
$t_1 = (p_{\ell} + p_{\nu})^2$ when the photon is radiated 
from the pion,
and by $t_2 = (p_K - p_{\pi})^2$ when the photon
is radiated from the charged lepton.
Note that $t_1 =  t_2$ without radiation,
but differ when radiation is included.
The $t$-dependence of the form factors is obtained from our 
measurements in~\cite{prd_ff}.

In the {\sc klor} program, the virtual and inner-\brems\
matrix elements are evaluated numerically.
Numerical integration of the squared matrix elements over 
the phase space gives predictions for \radcor, and also for the
radiative branching fractions, \rklrad.  
The virtual matrix elements are integrated over
three-body phase space,
and the IB matrix elements are integrated over 
four-body phase space.
Comparison of the predicted
and measured radiative branching fractions are presented in
Sec.~\ref{sec:results}.

 \section{Detector and Data Collection}
 \label{sec:apparatus}

An 800~GeV proton beam incident on a beryllium-oxide target produces
neutral kaons. A collimation system results in two parallel
neutral beams beginning 90~meters from the beryllium target;
each beam consists of roughly equal numbers of kaons and neutrons.
The fiducial decay region is 123-158 meters from the target,
and the vacuum region extends from 20-159~meters.

\begin{figure*}
  \centering
  \epsfig{file=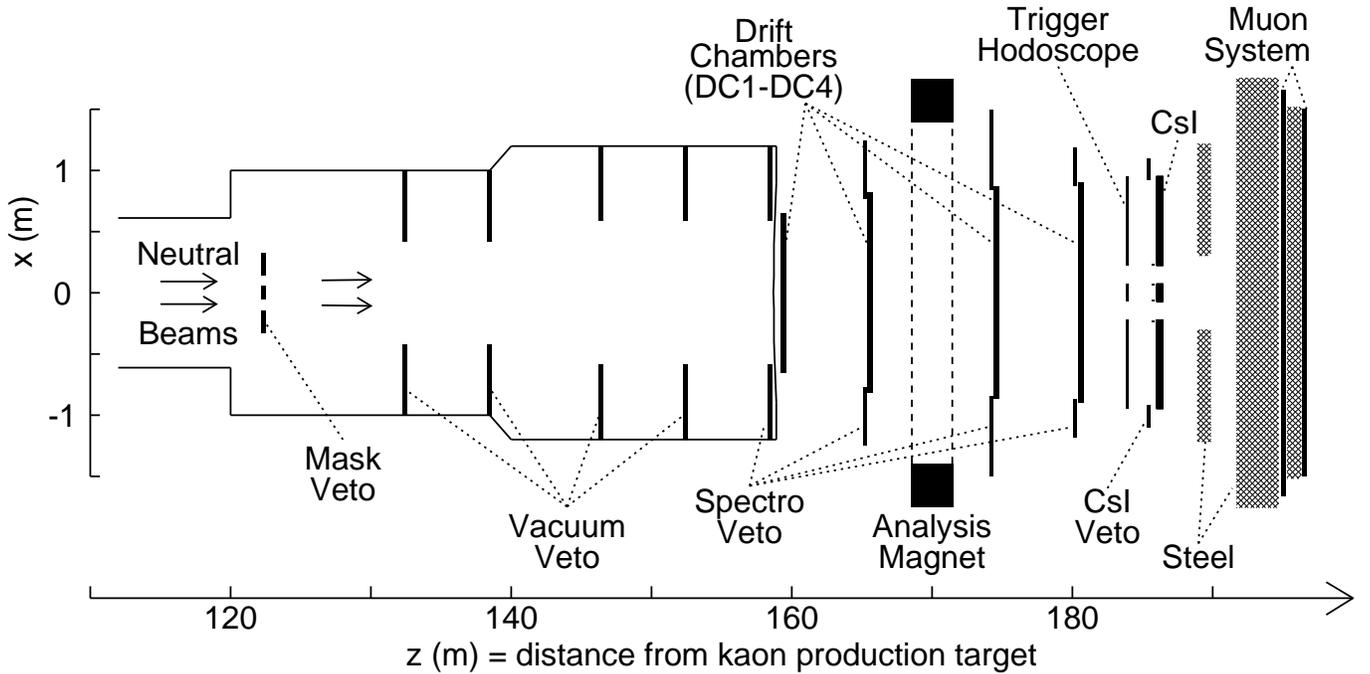,width=\linewidth}
  \caption{
    Plan view of the KTeV (E832) detector.
    The evacuated decay volume ends with a thin vacuum window at
    $Z = 159$~m.  $K_L$ decays from the two neutral beams 
    are the source of semileptonic decays.
           }
  \label{fig:detector}
\end{figure*}

\begin{figure}
  \centering
  \epsfig{file=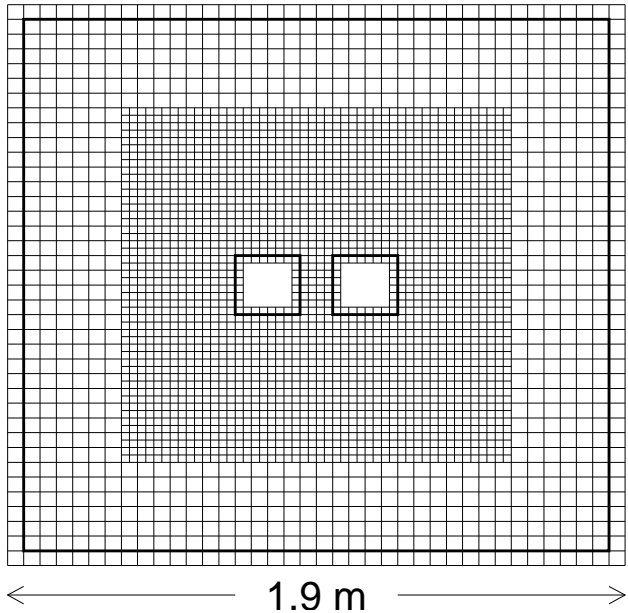,width=\linewidth}
  \caption{
       Layout of CsI calorimeter.
       The small crystals (inner region) have dimension
       $2.5\times 2.5\times 50~{\rm cm}^3$ ;
       the large crystals (outer region) have dimension
       $5.0\times 5.0\times 50~{\rm cm}^3$.       
       The two neutral beams go through the beam holes
       (into page) shown by the two inner squares.
       The fiducial cut, indicated by the dark lines, 
       excludes radiated photons that hit a crystal
       near the beam holes or near the outer boundary 
          }
  \label{fig:csi}
\end{figure}

The \ktev\ detector (Fig.~\ref{fig:detector})
is located downstream of the decay region.
A spectrometer consisting of four drift chambers,
two upstream and two downstream of a dipole magnet, 
measures the momentum of charged particles;
the resolution is 
$\sigma_p/p \simeq [1.7 \oplus (p/14) ] \times 10^{-3}$,
where $p$ is the track momentum in $\upk$.
Downstream of the spectrometer lies a  scintillator 
``trigger hodoscope,'' which is used to trigger on
charged particles. 
Farther downstream there is an electromagnetic calorimeter
made of 3100 pure Cesium Iodide (CsI) crystals 
[see Fig.~(\ref{fig:csi})].
For photons and electrons,
the energy resolution is better than 1\%
and the position resolution is about 1~mm.
The CsI calorimeter has two holes to
allow the neutral beams to pass through without interacting.
Two scintillator ``muon hodoscopes,'' 
behind 4 and 5 meters of steel, are used to detect muons.
The downstream muon hodoscope consists of horizontal (MUH) 
and vertical (MUV) counters, each with 15~cm segmentation.
Eight photon-veto detectors along the decay region and spectrometer
reject events with escaping particles.
The trigger for both semileptonic decay modes
requires a few hits in the drift chambers upstream of the magnet,
and at least two hits in the trigger hodoscope.

A Monte Carlo simulation is used to determine the acceptance
for the \Klthree\ and \Klthreerad\ decay modes.
{\sc klor} is used to generate  semileptonic decays 
as discussed in  Sec.~\ref{sec:phenom}.
For the generation of inner-\brems\ with {\sc klor},
the photon energy cutoff in the kaon rest frame is 3.5~keV.
This cutoff results in a minimum lab photon energy of
a few hundred keV, which is below the energy for which
photons affect the detector acceptance.
The fraction of \Klthree\ decays with a photon above the 
3.5~keV cutoff is 21\% for \Kethree, and 3.7\% for \Kmuthree.

Each of the decay products (pion, lepton, photon)
is traced through the detector, including the effects of
multiple scattering, energy loss from ionization, 
\brems\ for electrons, $e^+e^-$ pair-production for photons,
secondary decays for pions, and pions that interact hadronically
in the detector.
The effects of accidental activity are 
included by overlaying events
from a trigger that recorded random activity in the detector 
that is proportional to the instantaneous intensity of the proton beam.

In addition to detailed tracing of the decay products,
the \ktev\ MC also treats the detector response in great detail,
particularly the drift chamber inefficiencies and
detector resolutions. 
A detailed discussion of the simulation is given in~\cite{ktev03_reepoe}.

The semileptonic samples presented here were collected during a
one-day run in 1997 in which the incident proton 
beam intensity was reduced to 10\% of the nominal intensity used to
measure $\epsrat$~\cite{ktev03_reepoe}. 
The regenerator, which was used to generate $K_S$ in the \epsrat\
measurement, 
was removed from the beamline resulting in two  $K_L$ beams.
Note that this low-intensity sample is the same as that used 
for the $\Klthree$ form-factor measurements \cite{prd_ff}.

   \section{Analysis}
   \label{sec:anal}

The analysis strategy is first to identify a \Klthree\ decay
(Sec.~\ref{subsec:kl3_sel}),
and then search for an extra photon in the CsI calorimeter
(Sec.~\ref{subsec:rad_sel}).
The photon selection is optimized to reduce background from
semileptonic events with a fake photon cluster,
and also from other $K_L$ decays (mainly $\KLpmz$)
which are misidentified as \Klthreerad.
Kinematic requirements on the pion-lepton system for
\Klthree\ candidates
are also applied to the pion-lepton-photon system
for \Klthreerad\ candidates.

   \subsection{\KLpilnu\ Selection}
   \label{subsec:kl3_sel}

The first step in the semileptonic decay reconstruction
is to identify the charged pion and lepton.
The spectrometer is used to find two charged tracks.
For each track, the momentum is required to be above
$8~\upk$, and a cluster in the CsI calorimeter is required
to be near the track-projection at the CsI.
If the cluster energy ($E$) divided by the track momentum($p$)
is greater then 0.92, the track is identified as an electron;
this cut retains 99.8\% of the electrons and rejects
99.5\% of the pions.
If the corresponding $E/p$ value is less then 0.85, 
and the track does not point to a hit muon counter,
the track is identified as a pion;
this cut retains 99.1\% of the pions and rejects
99.93\% of the electrons.
A muon track candidate must have momentum above $10~\upk$,
point near a hit counter in both MUH and MUV
\footnote{
  For the track projection to the
  $15\times 15~{\rm cm}^2$ hit region of MUH and MUV,
  the proximity requirement is based on the amount of multiple
  scattering and is momentum dependent.
 }, 
and be matched to a CsI cluster with energy less than 2~GeV
(five times the average energy deposit).
Events consisting of either ``$\pi\mu$'' or ``$\pi e$'' candidates
are selected. Note that $\pi e$ 
candidates are vetoed by activity in the muon hodoscopes.

After particle identification, but before kinematic requirements,
there is a small background from \KLpmz\ ($\Kthreepi$) decays in which
the $\pz$-photons do not fire a photon veto,
and a pion either decays or is mis-identified as an electron.
At this intermediate stage of the analysis,
the $\Kthreepi$ background is 0.5\% in the $\pi\mu$ ($\Kmuthree$) sample 
and 0.1\% in the $\pi e$ ($\Kethree$) sample.
When a radiated photon cluster is required in the CsI calorimeter 
(see below), the $\Kthreepi$ background increases significantly:
67\% background in the \Kmuthreerad\ sample,
and almost 10\% in the \Kethreerad\ sample.
To suppress \Kthreepi\ background,
a kinematic variable is used which distinguishes
decays with a missing $\pz$ from decays with a missing neutrino;
we require $\ppzkin < -0.00625$ 
\footnote{For \KLpmz\ decays, \ppzkin\ is the 
                 longitudinal \pz\ momentum-squared in a frame in which
                  the $\ppc$ momentum is orthogonal to the kaon momentum.
                  See \cite{prd_br} for more details.
               }
for the $\pi e$ sample, 
and $\ppzkin < -0.01$ for the $\pi\mu$ sample.

For reconstructed \Klthree\ decays with a missing neutrino,
there is a twofold ambiguity in 
determining the kaon energy ($E_K$).
Both $E_K$ solutions are required to satisfy
the 40-160~GeV range for which the energy spectrum 
is well measured
\footnote{
   If there were no $E_K$ requirement, or a requirement only on the
   most probable $E_K$ solution,
   the analysis would accept decays with kaon energy outside
   the well-measured 40-160~GeV range, and thereby degrade
   the acceptance determination.
     }.
For reconstructed decays in which the true kaon energy is 
between 40 and 160~GeV, 20\% of these events are rejected
by the requirement on both kaon energy solutions.

After these selection requirements, 
there are $\NKmuthree$ $\Kmuthree$ 
and  $\NKethree$ $\Kethree$ candidates.
The background in each \Klthree\ mode
is $\sim 10^{-4}$, and is ignored.

   \subsection{ $\KLpilnu\gamma$ Selection}
   \label{subsec:rad_sel}

After finding a pion and lepton track that satisfy kinematics
for either of the \Klthree\ decays, 
we search for a photon in the CsI calorimeter.
A candidate radiated photon from a $\KLpilnu\gamma$ decay 
is a single CsI cluster with energy above 3~GeV
and a transverse profile consistent with a 
photon (Appendix~\ref{app:radshape}).
To have a well-defined acceptance, the photon position must not
lie in a crystal adjacent to the beam-hole, and must also lie
away from the outer edge of the calorimeter 
(Fig.~\ref{fig:csi}).
To avoid overlapping clusters in the CsI,
the photon cluster is required to be at least 20~cm from the 
lepton and at least 40~cm from the pion.

As with \Klthree\ decays, \Klthreerad\ decays have
two kaon energy solutions, both of which
must pass the 40-160~GeV requirement. 
Similarly, both $\Egcm$ solutions must be above 10~MeV.
For both \Klthree\ and \Klthreerad, the square of the
reconstructed neutrino momentum is required to be greater than zero.

After finding a single photon cluster in the CsI,
there is background from $\KLpmz$ decays, and from 
\Klthree\ decays with a fake photon cluster;
methods to reduce these backgrounds are described below.
Kaon decays other than $\KLpmz$ contribute  negligible background.

  \subsubsection{Background from Misidentified $K_L$ Decays}

\begin{figure}
  \centering
  \epsfig{file=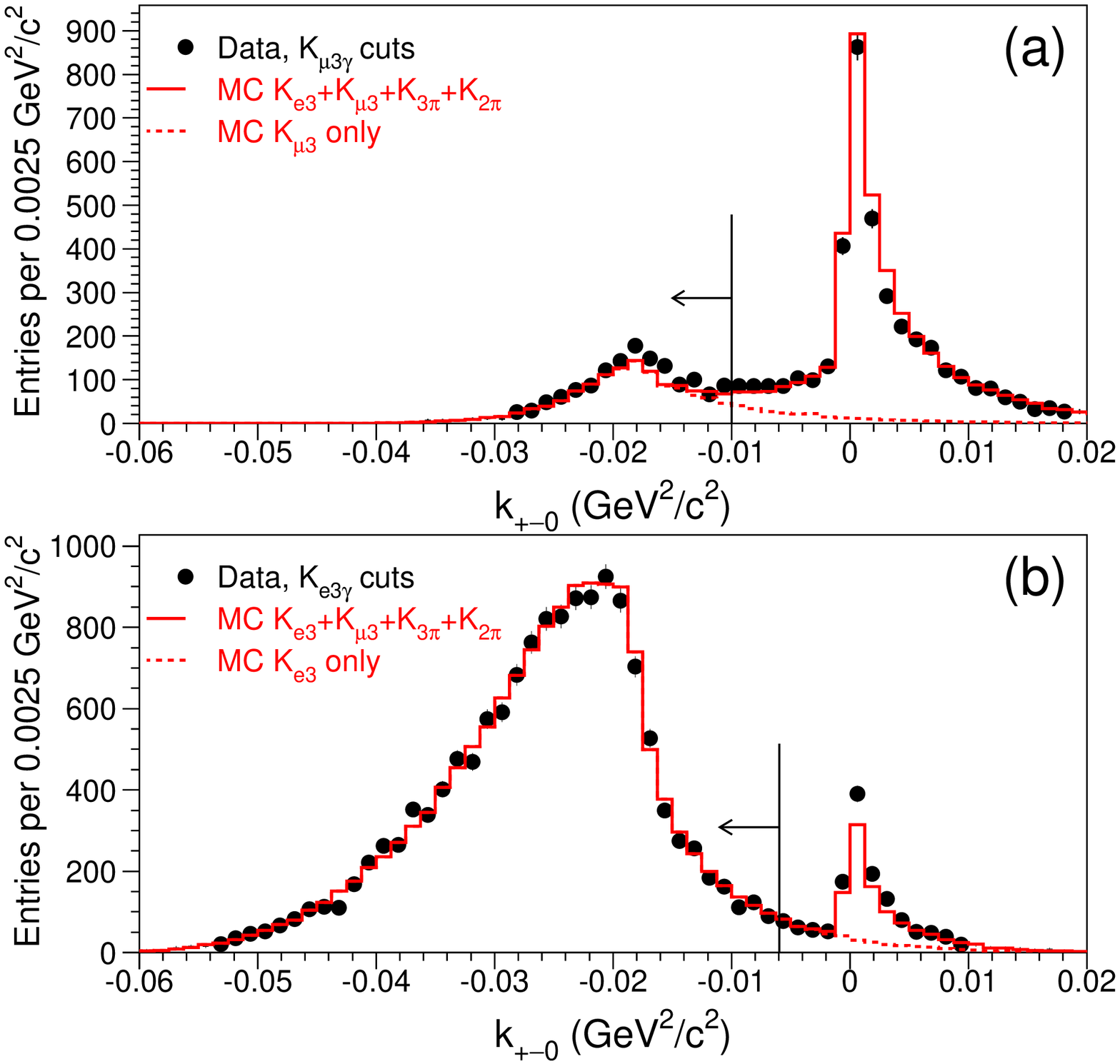,width=\linewidth}
  \caption{
      \ppzkin\ distribution for (a) \Kmuthreerad,
      and (b) \Kethreerad.  Arrows indicate
      selected events for the \Klthreerad\ samples.
      Each peak at $\ppzkin \sim 0$ is background from \KLpmz\ decays.
          }
  \label{fig:pp0kin}
\end{figure}

Background from \KLpmz\ decays arises when
one of the $\pz$ photons is not detected,
and a charged pion either decays or is misidentified as an electron.
As discussed in Sec.~\ref{subsec:kl3_sel},
this $\Kthreepi$ background is suppressed by the \ppzkin\ requirement.
Figure~\ref{fig:pp0kin} shows the \ppzkin\ distribution for 
both \Klthreerad\ samples; 
after the \ppzkin\ requirement,
the \Kthreepi\ background is $\KMUbgkpmz$\% in the \Kmuthreerad\ sample,
and less than 0.01\% in the \Kethreerad\ sample.
The 40\% relative uncertainty in the \Kthreepi\ background is
based on the data-MC discrepancy in the background region shown in
Fig.~\ref{fig:pp0kin}(b).

Backgrounds from other kaon decays ($K_{e4}$ and $\ppc\gamma$)
are simulated and normalized
according to their branching fractions.
Since there is no measurement or upper limit for $\KLpipimunu$,
we searched for this mode (Appendix~\ref{app:kmu4})
and set an upper limit on the
branching fraction of $B(\KLpipimunu) < 2\times 10^{-5}$.

  \subsubsection{Background from \Klthree\ With Fake Photon}

The other significant background is from 
\KLpilnu\ + ``fake-photon.''
Fake photon clusters can come from
(1) \pihad s in the detector,
(2) accidentals, and 
(3) external \brems\ upstream of the magnet.
For each fake-photon source,
the following subsections describe additional analysis requirements
to reduce these backgrounds;
background-to-signal ratios are determined with the MC.

{\it 1. Pion Hadron Interactions:}
If a {\pihad} leaves a fake photon candidate,
there are usually additional clusters in the CsI calorimeter
with transverse profiles inconsistent with a photon.
We therefore veto events that have an extra CsI cluster  
with energy above 1~GeV, and
a transverse profile inconsistent with a photon.
To avoid vetoing on satellite clusters from the pion shower
in the CsI, we veto an event only if the extra cluster
lies at least 30~cm away from the pion.
This extra cluster veto is required in both the \Klthree\ and
\Klthreerad\ samples.

The {\pihad} background is reduced further 
by requiring the photon cluster candidate
to lie at least 40~cm away from the pion at the CsI.
The background level is 
$\KMUbgpihad$\% for \Kmuthreerad, and
$\KEbgpihad$\% for \Kethreerad.
The uncertainty on this background is explained in
Appendix~\ref{app:pihad}.

\begin{figure}
  \centering
  \epsfig{file=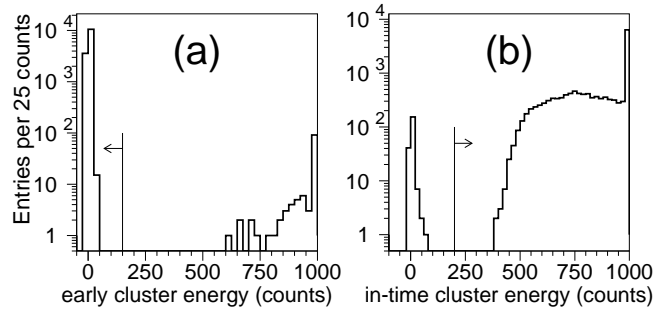,width=\linewidth}
  \caption{
      For radiated photon cluster candidates in the \Kethreerad\ sample,
      (a) cluster energy deposit in RF bucket prior to event, 
      and (b) cluster energy deposit in first RF bucket.
      Energy is shown as ADC counts (1~ADC count is about 1~MeV).
      The arrow in each plot 
      indicates the selected region.
          }
  \label{fig:csiadc}
\end{figure}

{\it 2. Accidental Cluster:}
To prevent an accidental cluster from faking a radiative photon,
we use the energy-vs-time profiles recorded by the CsI calorimeter.
For each radiated photon candidate, the CsI cluster energy deposited 
before the event must be consistent with pedestal, 
and the energy deposited in the first RF 
bucket~\footnote{
   The proton beam has a 53~MHz micro-structure such that protons
   (and hence neutral kaons) arrive in 1~ns wide RF-buckets,
   in 19~ns intervals. The CsI integration time is six RF buckets.
}
must be well above pedestal (Fig.~\ref{fig:csiadc}).
These two requirements on the energy-vs-time profile reduce the
accidental background by a factor of ten,
resulting in accidental background contributions of
$\KMUbgaccid$\% for \Kmuthreerad, 
and $\KEbgaccid$\% for \Kethreerad.
The relatively large error on the accidental background 
is because of the low statistics of the accidental event sample 
used to include accidental activity in the MC.

\begin{figure}
  \centering
  \epsfig{file=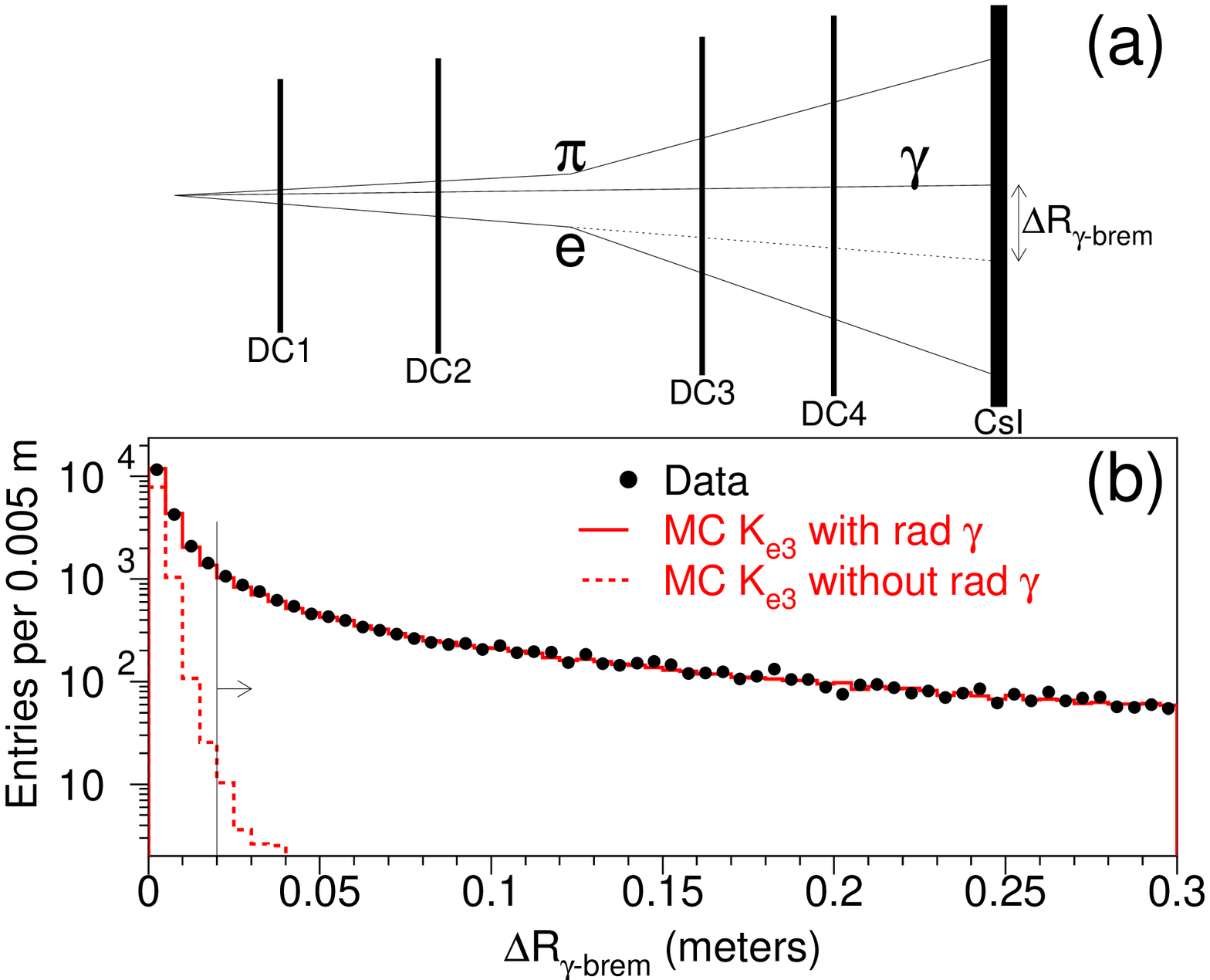,width=\linewidth}
  \caption{
      (a) Top view of spectrometer (DC1-4) and CsI calorimeter,
          and overlay of \Kethreerad\ decay.
      \drbrem\ is the distance between the photon candidate at the CsI 
      and the location of the electron track projected from 
      upstream of the magnet (dotted line);
      the latter is where a photon from external \brems\ would
      hit the CsI.
      (b) \drbrem\ distribution for
      data  (dots),  for MC with radiated photons (histogram)
      and for MC without radiated photons (dashed histogram).
      Plots are shown after
      \Kethreerad\ requirements.
      The arrow indicates the analysis selection.
          }
  \label{fig:xbrem}
\end{figure}

{\it 3. External \brems\ ($\Kethree$ only):}
A photon from external \brems\ upstream of the magnet
is separated from the electron at the CsI calorimeter,
and is therefore a source of fake photon clusters.
Note that external \brems\ downstream of the magnet
results in a photon that lies on top of the electron
at the CsI calorimeter, and therefore does not result in background.
To remove events with a photon produced by external \brems\
upstream of the magnet 
(which would fake $\Kethreerad$ with $\thgamecm \sim 0^0$),
the photon candidate is required to be at least 2~cm away from the
CsI position corresponding to the electron track projection from
upstream of the magnet [see \drbrem\ in Fig.~\ref{fig:xbrem}].
This background is $\KEbgxbrem$\%.

  \subsubsection{Background Summary}

\begin{figure}[hb]
  \centering
  \epsfig{file=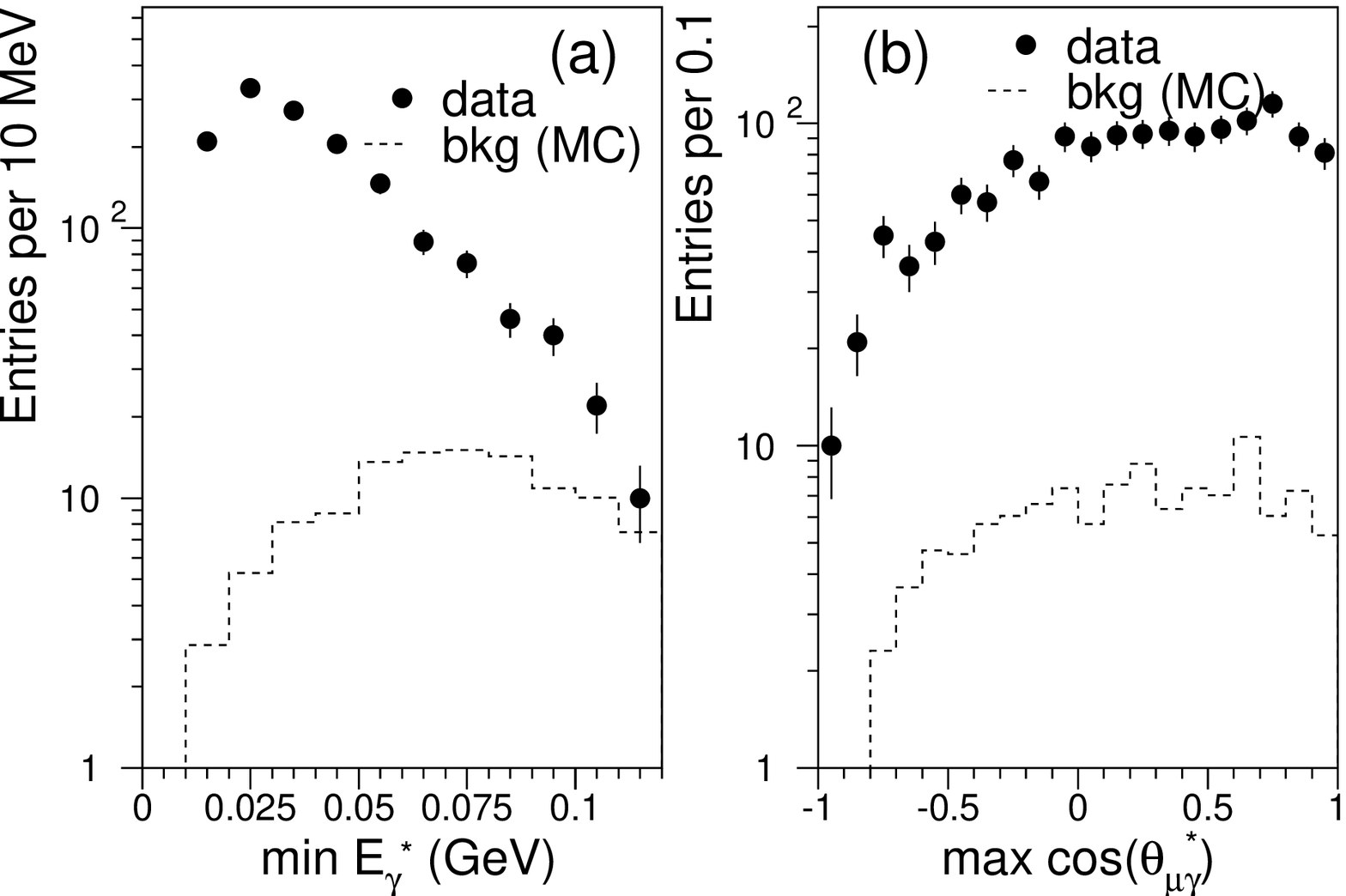,width=\linewidth}
  \caption{
     For $\KLpimunu\gamma$ candidates without background subtraction,
     (a) minimum $\Egcm$ and (b) maximum $\cos(\thgamecm)$ for the two
     kinematic solutions.
     Dots are data, and dashed histogram is the background 
     predicted by the MC.
          }
  \label{fig:bkgkm3cm}
\end{figure}

\begin{figure}[hb]
  \centering
  \epsfig{file=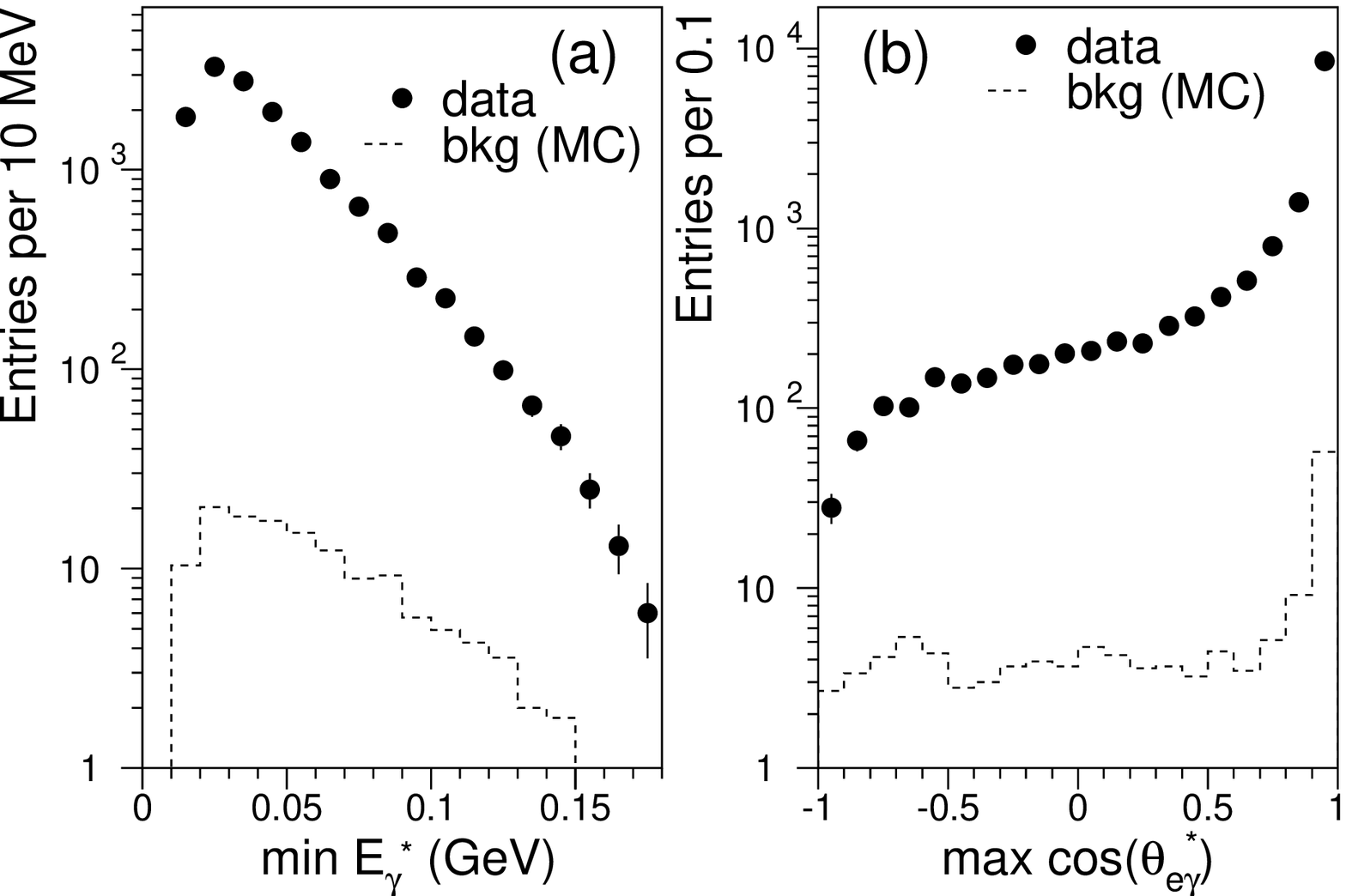,width=\linewidth}
  \caption{
     For $\KLpienu\gamma$ candidates without background subtraction,
     (a) minimum $\Egcm$ and (b) maximum $\cos(\thgamecm)$ for the two
     kinematic solutions.
     Dots are data, and dashed histogram is the background 
     predicted by the MC.
          }
  \label{fig:bkgke3cm}
\end{figure}

The backgrounds for both \Klthreerad\ modes are summarized
in Table~\ref{tb:bkg}.  The total background level is
$\KMUbgsum$\%   for $\Kmuthreerad$  and
$\KEbgsum$\%    for $\Kethreerad$.
Figures~\ref{fig:bkgkm3cm} and \ref{fig:bkgke3cm}
show \Egcm\ and $\cos(\thgamlepcm)$ distributions 
for the \Kmuthreerad\ and \Kethreerad\ decay modes, respectively.
In these plots, the data points correspond to the distributions
before background subtraction, 
and the dotted histograms show the background prediction.
The final \Klthreerad\ samples are obtained after subtracting
background coming from other kaon decays. 
The three $\Klthree +$ fake-photon backgrounds
are not subtracted from data since these effects are included
in the signal MC. 
After analysis selection and background-subtraction,
there are  
$\NKmuthreerad$ $\Kmuthreerad$ and 
$\NKethreerad$  $\Kethreerad$ candidates.


\begin{table}[hb]
  \centering
  \caption{
      \label{tb:bkg}
       The background-to-signal ratio (B/S) for each
       background component.
       }
  \medskip
\begin{ruledtabular}
\begin{tabular}{lccc}
    Background         & $\Kmuthreerad$   & $\Kethreerad$&   \\
    process            & B/S (\%)         & B/S (\%)         \\      
\hline 
  $\Klthree$ + \pihad\
          & $\KMUbgpihad$    & $\KEbgpihad$  \\
  $\Klthree$ + accidental cluster         
          & $\KMUbgaccid$    & $\KEbgaccid$  \\
  $\Klthree$ + external Brem 
          & ---   &  $\KEbgxbrem$  \\
  $\Kethree$ + particle mis-id      
          & $\KMUbgkethree$  &  ---    \\
  $\Kmuthree$ + particle mis-id     
          &     ---         &  $< 0.01$ \\
  $\KLpmz$ 
          & $\KMUbgkpmz$    & $ < 0.01 $  \\
  $\KLSpp\gamma$              
          & $\KMUbgkpmg$    & $ < 0.01 $  \\
  $\KLpipienu$      
          &      ---        &  $\KEbgke4$     \\
  $\KLpipimunu$      
          &     $\KMUbgkmu4$      &    ---     \\
\hline  
  Total   & $\KMUbgsum$\%   & $\KEbgsum$\%    \\
\hline  
\end{tabular}
\end{ruledtabular}
\end{table}

 \section{ Systematic Uncertainties }
 \label{sec:syst}

The $\Klthreerad/\Klthree$ ratios are insensitive to the
small inefficiencies related to track-finding
and particle identification.  
The main sources of systematic uncertainty are related
to \Klthreerad\ background,
the efficiency of reconstructing a radiated photon,
and the aperture for photons.
These systematic uncertainties are discussed below.

{\it Background:}
For $\rkmurad$, the background contributes $\KMUBGsyst$\%
uncertainty, mostly from \KLpmz\ decays;
for $\rkerad$  the corresponding uncertainty is $\KEBGsyst$\%,
mainly from \Kethree\ with a false photon cluster.

{\it Extra Cluster Veto: }
Removing the extra cluster veto requirement changes 
\rkmurad\ by $-\KMUXCLVTO$\% and changes 
\rkerad\ by $+\KEXCLVTO$\%; these changes are included
as uncertainties in $\rklrad$.

{\it CsI Energy Scale: }
The CsI energy scale is calibrated to better than 0.1\%, and results
in a 0.1\% uncertainty for both $\rklrad$  ratios.

{\it Photon Aperture and Beamline Material: }
The beam-hole aperture is studied with electrons from \Kethree\ decays.
The effective size is known to $\pm 200~\mu$m, leading
to a 0.1\% uncertainty for both $\rklrad$ ratios.
The same $200~\mu$m uncertainty affects the \drbrem\ requirement,
resulting in a 0.30\% uncertainty in \rkerad, 
and less than $0.01$\% uncertainty in \rkmurad.

{\it Photon Inefficiency: }
Three sources of photon inefficiency are considered:
$\gamma\to e^+e^-$ conversions in detector material,
CsI readout, and the transverse profile requirement.
A photon traverses 4\% of a radiation length through the
 \ktev\ detector before reaching the CsI calorimeter,
and this results in a 1\% reconstruction loss for the radiated photon.
Based on the data-MC agreement in the first two bins of
Fig.~\ref{fig:xbrem}, the amount of material is known to 
better than 10\%, resulting
in a systematic uncertainty of 0.1\% on $\rklrad$.
The CsI readout efficiency is studied with events
in which light from a laser is distributed to each crystal via
3100 fibers,
and muons are used to check dead material (wrapping) 
between the crystals; this results in a 0.01\% uncertainty.
The inefficiency of the transverse profile requirement 
is discussed in Appendix~\ref{app:radshape}, resulting in a
systematic uncertainty of 0.2\% for \Kethreerad,
and 0.06\% for \Kmuthreerad.

{\it Muon Hodoscope Efficiency:} 
In the MC,
if muon scattering in the steel is turned off
and gaps between the muon system counters are ignored,
\rkmurad\ changes by 2\%.
Using the nominal MC that includes scattering and gaps,
we compare data and MC distributions of the distance
between the track extrapolation and the hit muon system counters;
this comparison shows that muon propagation through the steel
is described to better than 20\% precision, leading to a 
systematic uncertainty of 0.4\% on \rkmurad.

{\it Radiative Effects and Form Factors in Matrix Element: }
As discussed in Sec~\ref{subsec:rad2}, the uncertainty in 
radiative effects is from second-order radiative correction; 
this leads to \RADRADERRPERCENT\% uncertainty in \rklrad.
Variations in the form factors measured by \ktev~\cite{prd_ff} 
lead to 0.1\% uncertainty on \rklrad.

\bigskip
The uncertainties are summarized in Table~\ref{tb:syst}. 
The total systematic uncertainty is
$\RKMURADrelsyst$\% for $\rkmurad$, and
$\RKERADrelsyst$\% for $\rkerad$.
These systematic uncertainties are comparable to
the statistical uncertainties.

\begin{table}[hb]
  \centering
  \caption{
      \label{tb:syst}
       Summary of systematic and statistical uncertainties.
       The total uncertainty in the last row is the sum (in quadrature)
       of the systematic and statistical uncertainties.
       }
  \medskip
\begin{ruledtabular}
\begin{tabular}{l | cc}
   Source of          & \multicolumn{2}{c}{\% Uncertainty for:} \\
   Uncertainty        & $\rkmurad$   & $\rkerad$         \\ 
\hline 
  background                   &   $\KMUBGsyst$ &  $\KEBGsyst$   \\
  extra cluster veto           &   $\KMUXCLVTO$ &  $\KEXCLVTO$   \\
  CsI energy scale             &    0.10        &     0.08        \\
  $\gamma$ beam-hole aperture  &    0.09        &     0.09       \\
  $\drbrem$                    &    $<0.01$     &     0.30       \\
  photon inefficiency          &    0.12        &     0.22       \\
  muon hodoscope inefficiency  &    0.4         &     --         \\
  form factors                 &    0.1         &     0.1        \\
  2nd order rad corrections    & $\RADRADERRPERCENT$ & $\RADRADERRPERCENT$ \\
  MC signal statistics         & $\RKMURADmcstat$ & $\RKERADmcstat$  \\
\hline  
  Total Systematic          & $\RKMURADrelsyst$\%   & $\RKERADrelsyst$\%
   \\
  Data Statistics           &  $\RKMURADrelstat$\%  & $\RKERADrelstat$\%
   \\
  Total Uncertainty         &   $\RKMURADreltot$\%  & $\RKERADreltot$\%
\end{tabular}
\end{ruledtabular}
\end{table}

 \section{Extraction of Results}
 \label{sec:extraction}

  \subsection{Acceptance Corrections}
  \label{subsec:acc}

The relative branching fraction for radiative semileptonic decays
is determined from the data and MC acceptance corrections by
\begin{equation}
   \rklrad = \frac{ N^{data}_{K\ell 3\gamma}/A_{K\ell 3\gamma} }
                  { N^{data}_{K\ell 3}/A_{K\ell 3} } ~,
\end{equation}
where $\NDATA$ is the number of reconstructed decays
in data, and $A_{mode}$ 
is the acceptance determined by the 
Monte Carlo simulation. 
Note that the subscript $mode$ refers to either 
$\Klthree$ or $\Klthreerad$.
The acceptance for each mode is defined as
\begin{equation}
    A_{mode} = \NMCREC / \NMCGEN
    \label{eq:accdef}
\end{equation}
where $\NMCGEN$ and $\NMCREC$ are the numbers of generated
and reconstructed events, respectively.
$\NMCGENKL$ is the number of generated $\KLpilnu(\gamma)$ events in which
the generated kaon energy and decay vertex are within the
nominal ranges (40-160~GeV$/c$ and 123-158~m).
$\NMCGENKLg$ is the number of generated $\KLpilnu\gamma$ events
with a radiated photon energy greater than 10~MeV,
and with the same energy and vertex requirements as for $\NMCGENKL$.
For each sample of generated \Klthree\ and \Klthreerad\ events,
$\NMCRECKL$ and $\NMCRECKLg$ are the numbers of events that
satisfy the selection requirements in Sec.~\ref{sec:anal}.
The approximate acceptance is 20\% for \Klthree\ decays
and 3\% for \Klthreerad\ decays.
For \Kethreerad, the acceptance is corrected for
second-order radiative effects as described below.

  \subsection{Second-Order Radiative Corrections}
  \label{subsec:rad2}

First-order radiative effects in \Klthree\ decays are fully treated as
described in Sec.~\ref{sec:phenom}.
The precision of this analysis, however, 
requires a correction from second-order radiative effects
involving two radiated photons.
With two radiated photons, the definition of \rklrad\
is modified to include semileptonic decays with at least
one photon above 10~MeV.
To see why second-order radiative effects are needed,
consider the \KLpienu\ acceptance. The MC with first-order 
radiative effects changes the acceptance by 2.5\% compared
to the Born-level (zero order) MC without radiative effects. 
This correction suggests that to determine the acceptance for 
radiative $\KLpienu\gamma$ (first order),
there is a few percent acceptance correction from
the next (second) order in radiative corrections.
Note that this dependence on higher order radiative effects
depends on the experimental apparatus.
The KTeV sensitivity is largely from the kinematic selection
and the hermetic veto system with low thresholds (150~MeV).

The second-order radiative correction to the acceptance 
is estimated using {\sc photos}~\cite{photos2},
which allows for up to two radiated photons.
We begin by generating MC with single-photon radiation from {\sc photos},
and then reweight {\sc photos} to match data distributions
in the variables $\Egcm$ and $\thgamlepcm$
\footnote{{\sc photos} without reweighting agrees
           with data in the $\Egcm$ variable, 
           but has a large disagreement in the \thgamlepcm\ 
           distribution. See Appendix~C of~\cite{Troy}
           for discussion of {\sc photos}.
              }.
For two-photon generation with {\sc photos}, 
the reweighting is applied to the higher energy photon.
The correction to the radiative acceptance
for second-order effects is given by
\begin{equation}
  \RADRAD \equiv A_{{\rm PHOTOS}-2\gamma} / A_{{\rm PHOTOS}-1\gamma}~,
\end{equation}
where $A_{{\rm PHOTOS}-1,2\gamma}$ is the acceptance determined
with {\sc photos} using the option to allow up to 1 or 2 
radiated photons.
The final acceptance ($A_{K\ell 3\gamma}$)
is given by \RADRAD\ times the acceptance determined from the 
MC using {\sc klor}~\cite{Troy}.

Evidence for a second radiated photon is seen in the
photon candidate multiplicity in the CsI calorimeter.
The photon multiplicity is checked for \Klthree\ candidates
after requiring all photon identification requirements
(cluster energy greater than 3~GeV, transverse shape, etc.),
but before \Klthreerad\ kinematics are applied.
The fraction of these events with two photon candidates ($\ETWO$)
is $\ETWO = (0.39\pm 0.05)$\% in data.
In our MC using the {\sc klor} generator,
events with two photon candidates are mostly from accidentals,
and $\ETWO = (0.17\pm 0.01)\%$.
For MC using {\sc photos} with one radiated photon,
$\ETWO = (0.18\pm 0.01)$\%.
These MC generators ({\sc klor} and {\sc photos}) agree with each other, 
but disagree with the data.
For MC using  {\sc photos} with two radiated  photons,
$\ETWO = (0.36\pm 0.02)$\%;
this good agreement with data shows that {\sc photos} provides
a reasonable second-order correction to the acceptance.

For $\KLpienu\gamma$, 
$\RADRAD = \RADRADVALUE \pm \RADRADERRABS$
\footnote{
    This 1.8\% second-order acceptance correction is much 
    larger than the 0.2\% probability of observing a second 
    radiative photon in the CsI calorimeter.
    The 3~GeV minimum cluster energy requirement severely limits the
    chance of observing a second radiated photon, 
    while radiated photons with much less energy effect the acceptance.    
},
where the uncertainty comes from the limited statistics
in the data-MC comparison of events with two photon 
candidates.
For $\KLpimunu\gamma$, the two-photon candidates are consistent
with accidentals, and \RADRAD\ is consistent with one.
We therefore set $\RADRAD = 1.0$ for $\KLpimunu$, 
but include an additional \RADRADERRPERCENT\% uncertainty on $\rkmurad$.

 \section{ Results }
 \label{sec:results}

  \subsection{ $\KLpimunu\gamma$ Branching Fraction }
  \label{subsec:km3rad_result}

\begin{figure}[hb]
  \centering
  \epsfig{file=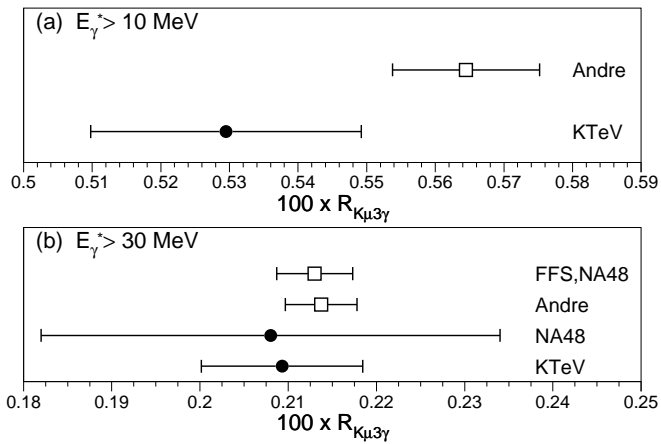,width=\linewidth}
  \caption{
      Comparison of \rkmurad\ among experiment and theory
      for (a) $\Egcm > 10$~MeV, and (b) $\Egcm > 30$~MeV.
      The experimental results are from \ktev\ and 
      NA48~\cite{na48_kmu3rad}.
      The theory predictions are from Andre~\cite{Troy}
      and from Fearing, Fischbach, and Smith, FFS~\cite{FFS}.
      The FFS prediction is determined by NA48, and is corrected by 
      $(1+\delta_K^{\mu})^{-1}$~\cite{Troy}.
      The theory uncertainties
      are taken to be $\delta_K^{\mu} = 0.019$ times the prediction.
          }
  \label{fig:brkm3rad_compare}
\end{figure}

For the muonic semileptonic decay mode,
the radiative branching fraction is 
\begin{eqnarray}
    \rkmurad(\Egcm & > & 10~{\rm MeV})  = 
                      \nonumber  \\ 
            & = &
        [\RKMURAD \pm \RKMURADerrstat(stat) \pm \RKMURADerrsyst(syst)]\%
                        \nonumber \\
            & = &
        [\RKMURAD \pm \RKMURADerrtot ]\% ~.
\end{eqnarray}
The only previous measurement of this radiative mode 
\cite{na48_kmu3rad}
required $\Egcm > 30$~MeV.
To compare with this measurement, we have also applied a 30~MeV
requirement on the photon energy;
the results are
\begin{eqnarray}
   \rkmurad({\rm KTeV};~\Egcm > 30~{\rm MeV}) 
           & = &  [\OLDKMURAD \pm \OLDKMURADerrtot]\%   \\
   \rkmurad({\rm NA48};~\Egcm > 30~{\rm MeV}) 
           & = &  [0.208\pm 0.026]\%         
\end{eqnarray}
which are in good agreement. The \ktev\ result is almost three times
more precise than the previous measurement.
Figure~\ref{fig:brkm3rad_compare} compares our measurement
of \rkmurad\ with NA48, and with the predictions from
Andre~\cite{Troy} and FFS~\cite{FFS}. These comparisons
show good agreement between the measurements and theory predictions.
Table~\ref{tb:summkm3} summarizes the \rkmurad\ results from \ktev\
and the predictions from {\sc klor}.

\begin{table}[hb]
  \centering
  \caption{
      \label{tb:summkm3}
  Summary of \rkmurad\ values measured by \ktev\ and 
  predicted by {\sc klor}.
       }
  \medskip
\begin{ruledtabular}
\begin{tabular}{cc | ccc}
  min     & min          &       &        &  \\
  $\Egcm$ & $\thgammucm$ & KTeV  & KLOR   & KTeV/KLOR \\
   (MeV)  &   (deg)  &
  result (\%) & prediction (\%) &   ratio  \\
\hline 
    10  &  0  & $\RKMURAD \pm \RKMURADerrtot$      
              & $\KLORKMRAD$
              & $\KMRADRATIO$  \\
    30  &  0  & $\OLDKMURAD \pm \OLDKMURADerrtot$  
              & $\KLOROLDKMRAD$
              & $\OLDKMRADRATIO$  \\
\end{tabular}
\end{ruledtabular}
\end{table}

  \subsection{ $\KLpienu\gamma$ Branching Fraction }
  \label{subsec:ke3rad_result}

\begin{figure}[hb]
  \centering
  \epsfig{file=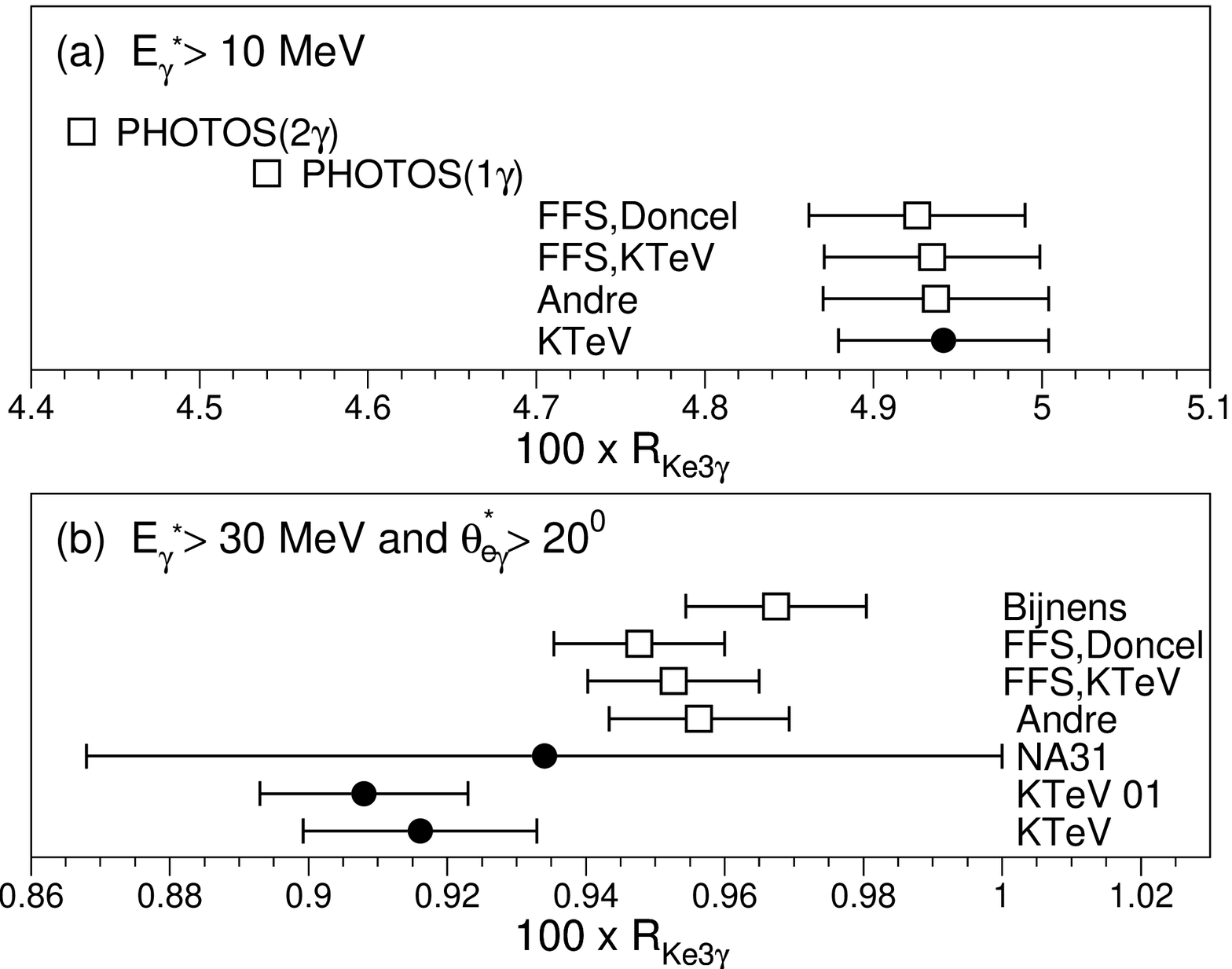,width=\linewidth}
  \caption{
      Comparison of \rkerad\ among experiment and theory
      for (a) $\Egcm > 10$~MeV, and 
      (b) $\Egcm > 30$~MeV and $\thgamecm > 20^0$.
      The experimental results are from \ktev\ 
      (this result and also \ktev\ 01~\cite{ktev01_ke3rad}),
      and NA31~\cite{na31_ke3rad}.      
      The theory predictions are from
      Andre~\cite{Troy},
      Fearing, Fischbach, and Smith, FFS~\cite{FFS}, 
      Bijnens~\cite{Bijnens93}, and {\sc photos}~\cite{photos2}.
      The FFS prediction is determined by
      Doncel~\cite{Doncel} and by \ktev,
      and each is corrected by 
      $(1+\delta_K^{e})^{-1}$~\cite{Troy}.
      The theory uncertainties are taken to be 
      $\delta_K^e = 0.013$ times the prediction.
      The {\sc photos} predictions for one and two radiative
      photons are described in Sec.~\ref{subsec:rad2}.
          }
  \label{fig:brke3rad_compare}
\end{figure}

For the electronic semileptonic decay mode, the
branching fraction is
\begin{eqnarray}
   \rkerad(\Egcm & > & 10~{\rm MeV})  = 
                        \nonumber \\
           & = &
        [\RKERAD \pm \RKERADerrstat(stat) \pm \RKERADerrsyst(syst)]\%
                        \nonumber \\
            & = &
        [\RKERAD \pm \RKERADerrtot ]\% ~.
\end{eqnarray}
Figure~\ref{fig:brke3rad_compare}(a) compares our measurement 
of \rkerad\ with NA31, and with several theory predictions.
The measurements and predictions show good agreement,
except for the {\sc photos} prediction.
Although the {\sc photos} prediction is not consistent with
the measurements, it provides the only estimate
on the impact of second-order radiative effects;
{\sc photos} suggests that second-order
effects lower the predicted value of \rkerad\ by a few percent.

Previous measurements \cite{na31_ke3rad,ktev01_ke3rad}
were done with $\Egcm > 30$~MeV and $\thgamecm > 20^0$.
To compare with these measurements, we apply the same
energy and angle requirements; the results are
\begin{eqnarray}
 &  \rkerad(  {\rm NA31~96};  & ~\Egcm > 30~{\rm MeV},~\thgamecm > 20^0) 
      \nonumber \\     & = &  (0.934 \pm 0.066)\%    
                \label{eq:na31} \\
  & \rkerad(  {\rm KTeV~01};  & ~\Egcm > 30~{\rm MeV},~\thgamecm > 20^0) 
      \nonumber \\     & = &  (0.908 \pm 0.015)\%    
                  \label{eq:ktev01} \\
   & \rkerad(  {\rm KTeV~04};  & ~\Egcm > 30~{\rm MeV},~\thgamecm > 20^0) 
      \nonumber \\     & = &  [\OLDKERAD \pm \OLDKERADerrtot]\%  
         \label{eq:ktev04}~,
\end{eqnarray}
and are consistent with each other.
The current \ktev\ result is based on 4309 candidates, which is less
than 1/3  of the \ktev\ 01 sample
\footnote
   { The \Kethreerad\ selection requirements for the 
     \ktev~01 sample \cite{ktev01_ke3rad} were significantly 
     more relaxed compared to the current analysis, 
     and therefore resulted in  higher statistics.
   }.
The loss in statistical precision is compensated by a significant
reduction in the systematic uncertainty. 
The new analysis reported here 
(with $\Egcm > 30~{\rm MeV}$ and $\thgamecm > 20^0$)
therefore has comparable precision with our earlier result,
yet these two \ktev\ results are largely uncorrelated.
The new \ktev\ result (Eq.~(\ref{eq:ktev04})) supersedes the 
previous result~\cite{ktev01_ke3rad}
so that our radiative branching fractions
are all based on the same analysis method.
Figure~\ref{fig:brke3rad_compare}(b) compares the measurements
in Eqs~(\ref{eq:na31}--\ref{eq:ktev04})
with theory predictions.
Table~\ref{tb:summke3} summarizes the \rkerad\ results from \ktev\
and the predictions from {\sc klor}.

\begin{table}[hb]
  \centering
  \caption{
      \label{tb:summke3}
  Summary of \rkerad\ values measured by \ktev\ and 
  predicted by {\sc klor}.
       }
  \medskip
\begin{ruledtabular}
\begin{tabular}{ cc | ccc}
  min     & min          &       &        &  \\
  $\Egcm$ & $\thgamecm$& KTeV  & KLOR   & KTeV/KLOR \\
   (MeV)  &   (deg)    &
  result (\%) & prediction (\%) &   ratio  \\
\hline 
   10  &  0   & $\RKERAD \pm \RKERADerrtot$      
              & $\KLORKERAD$
              & $\KERADRATIO$  \\
   30  &  20  & $\OLDKERAD \pm \OLDKERADerrtot$  
              & $\KLOROLDKERAD$
              & $\OLDKERADRATIO$  \\
\end{tabular}
\end{ruledtabular}
\end{table}

  \subsection{Discussion}
  \label{subsec:discuss}

Our treatment of radiative effects propagates 
into the detector acceptance determination
for the \KLpilnu\ branching fractions~\cite{prd_br}
and form factors~\cite{prd_ff}.
The uncertainty in the acceptance from radiative effects is based
on the agreement between our measurements of $\rklrad$
and the predictions from {\sc klor}~\cite{Troy}.
Specifically, this uncertainty is taken to be the 
measurement-prediction difference, 
plus the quadrature-sum of the experimental and theoretical
uncertainties
(from Tables~\ref{tb:summkm3} and \ref{tb:summke3} ).

For $\KLpimunu\gamma$, with $\Egcm > 10$~MeV, 
the uncertainty in our treatment of radiative effects 
is evaluated to be 10.1\%;
with $\Egcm > 30$~MeV, this uncertainty is 6.7\%.
For $\KLpienu\gamma$, with $\Egcm>10$~MeV, 
the corresponding uncertainty is 1.9\%;
with $\Egcm > 30$~MeV and $\thgamecm > 20^0$, 
the uncertainty is 6.4\%.
From these data-theory comparisons, which are all consistent,
we assign the average uncertainty of 6\% on {\sc klor}'s
treatment of radiative effects.

\begin{figure}[hb]
  \centering
  \epsfig{file=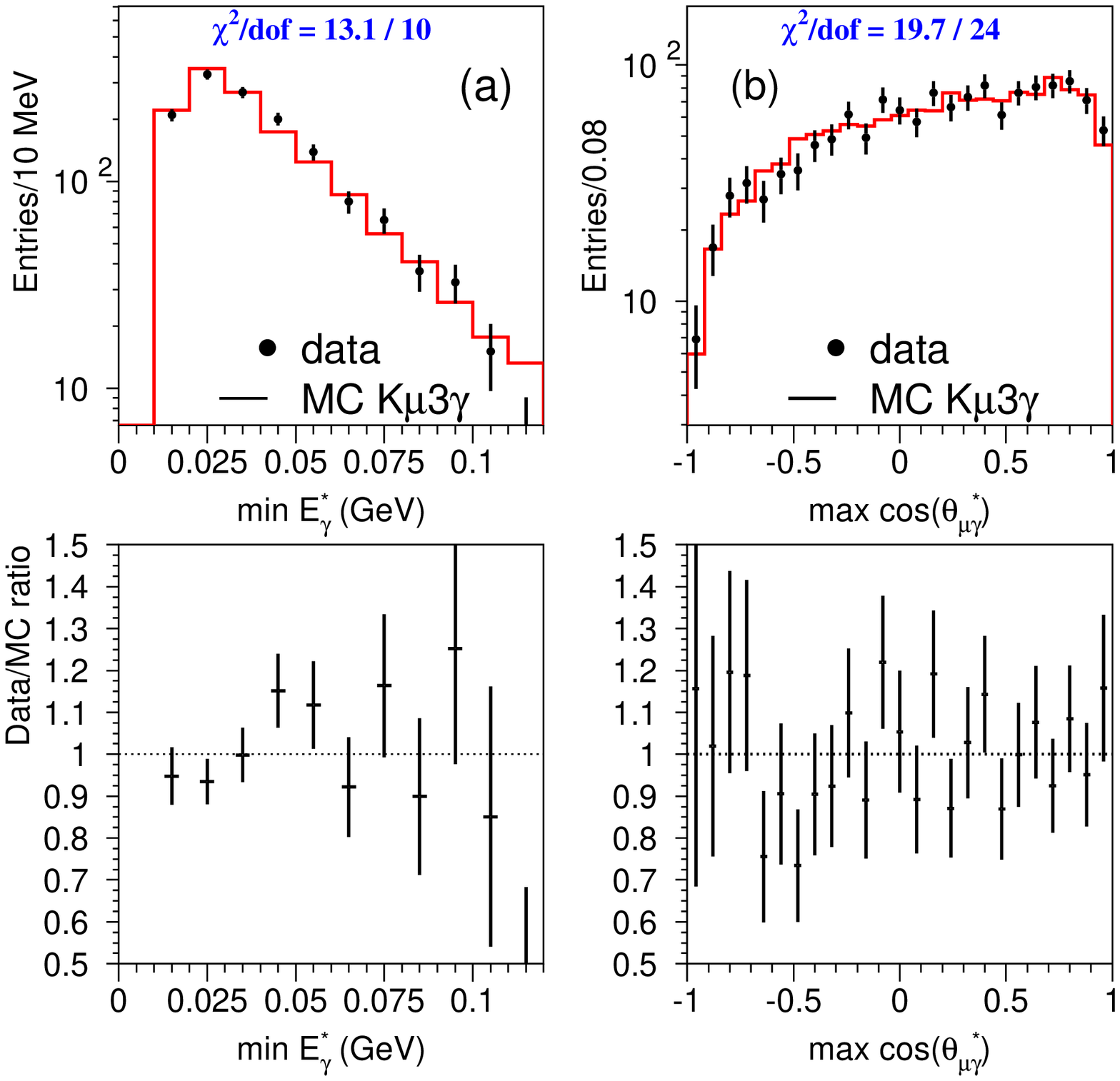,width=\linewidth}
  \caption{
      Data-MC comparison of $\Kmuthreerad$ kinematics.
      For the two kaon energy solutions in the lab,
      (a) minimum radiated photon energy in kaon rest frame, and
      (b) maximum cosine of the photon-muon angle in kaon rest frame.
      Data are shown in dots; MC in histogram.
      The $\chi^2/dof$ at the top of each plot refers to the
      data-MC comparison; data/MC ratios are shown in the lower plots.
      All $\Kmuthreerad$ selection requirements have been applied,
      and background is subtracted.
          }
  \label{fig:km3kin}
\end{figure}

\begin{figure}[hb]
  \centering
  \epsfig{file=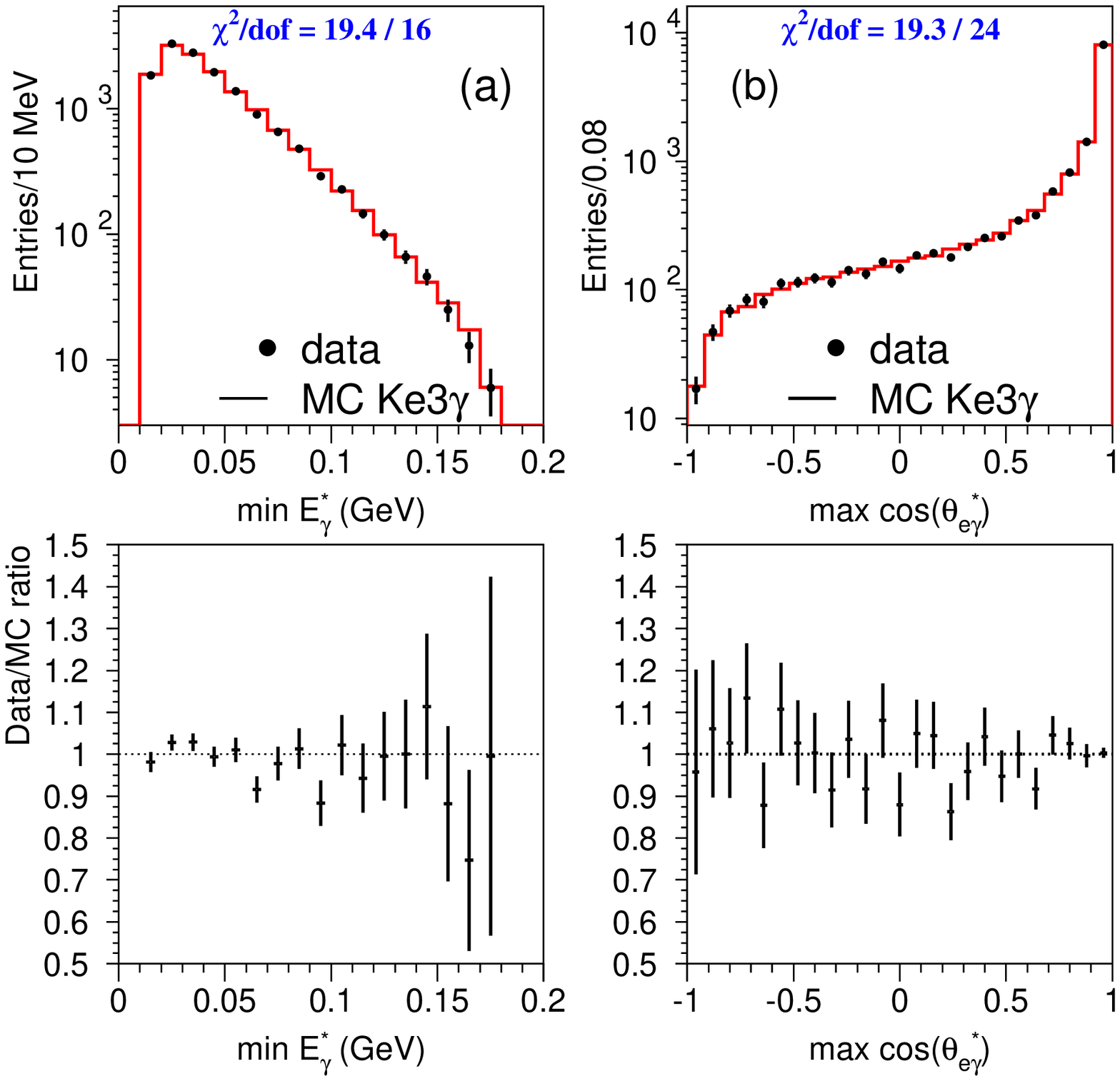,width=\linewidth}
  \caption{
      Data-MC comparison of $\Kethreerad$ kinematics.
      For the two kaon energy solutions in the lab,
      (a) minimum radiated photon energy in kaon rest frame, and
      (b) maximum cosine of the photon-electron angle in kaon rest frame.
      Data are shown in dots; MC in histogram.
      The $\chi^2/dof$ at the top of each plot refers to the
      data-MC comparison; data/MC ratios are shown in the lower plots.
      All $\Kethreerad$ selection requirements have been applied.
          }
  \label{fig:ke3kin}
\end{figure}

The generated phase space distributions from {\sc klor} 
are important in the form factor measurements~\cite{prd_ff},
and also in the $\rklrad$ measurements (Eq.~(\ref{eq:rkpilnurad})).
The {\sc klor} generator is checked by comparing
data and MC distributions for the photon energy ($\Egcm$)
and for the angle between the photon and charged lepton ($\thgamlepcm$).
Figures~\ref{fig:km3kin} and \ref{fig:ke3kin} show these data-MC 
comparisons for \Kmuthreerad\ and \Kethreerad, respectively.
Data and MC agree well in all cases, as indicated by the
$\chi^2/dof$.  
The data-MC comparison of the pion-photon angle (not shown)
also agrees well for both radiative decay modes.
The good quality of these data-MC comparisons shows that
the {\sc klor} generator is adequate for our experimental
sensitivity.

  \section{Conclusion}
  \label{sec:conclude}


For the radiative semiletponic decays
$\KLpimunu\gamma$ and $\KLpienu\gamma$,
we have measured the branching fractions 
and the distributions of photon energy
and photon-lepton angle (in the kaon center-of-mass frame).
The results are consistent with predictions based
on the {\sc klor} program~\cite{Troy}.
Our radiative branching fractions are also consistent with
previous measurements~\cite{na31_ke3rad,ktev01_ke3rad,na48_kmu3rad},
and with predictions based on the model of
Fearing, Fischbach, and Smith~\cite{FFS}.

\clearpage
\appendix


  \section{Efficiency of the Photon Transverse Profile Requirement}
  \label{app:radshape}

The transverse profile for energy deposits in the CsI calorimeter
is based on the energy distribution in the
crystals that are used to determine the cluster energy.
A CsI cluster uses 49 small crystals 
($2.5\times 2.5\times 50~{\rm cm}^3)$ or
9 large crystals
($5.0\times 5.0\times 50~{\rm cm}^3)$.
To determine the photon likelihood for a cluster,
we use a \schi\ variable,
\begin{equation}
   \schi \equiv \sum_{i=1}^{N} 
           \frac{ (f_i^{data} - \bar{f}_i)^2 }{ {\rm RMS}_i^2 }
\end{equation}
where $f_i^{data}$ is the fraction of energy in the $i$'th crystal,
$\bar{f}_i$ is the 
position-dependent average energy fraction in the  $i$'th crystal
as measured in \KLzz\ decays, and RMS$_i$ is the measured RMS
of $\bar{f}_i$. To reduce background in the \Klthreerad\ analyses,
we require $\schi < 10$, a considerably stricter requirement
than used in our other analyses involving 
\KLzz\ and $\KLzzz$ decays~\cite{ktev03_reepoe,prd_br}.

In the $\KLpienu\gamma$ analysis,
the inefficiency of the \schi\ requirement for photons is 2.0\%
as determined from MC. Most of the loss is from 
soft external \brems\ photons that overlap the 
radiated photon.
Contributions to the \schi\ tail are understood with 10\% precision,
resulting in 0.2\% uncertainty on \rkerad.
For $\KLpimunu\gamma$, the photon inefficiency is 0.6\%,
resulting in a 0.06\% uncertainty in \rkmurad.

As a crosscheck in the $\KLpienu\gamma$ analysis, 
we consider a stricter requirement of 
$\schi < 3$; the additional loss is $(5.6\pm 0.2)$\% in data,
and $(5.8\pm 0.1)$\% in MC.

\def\km4{ K_L \to \pi^0 \pi^{\pm} \mu^{\mp} \nu }
\def\mmunu{m_{\mu\nu}}
  \section{Search for $\km4$\ }
  \label{app:kmu4}

\begin{figure}[hb]
  \centering
  \epsfig{file=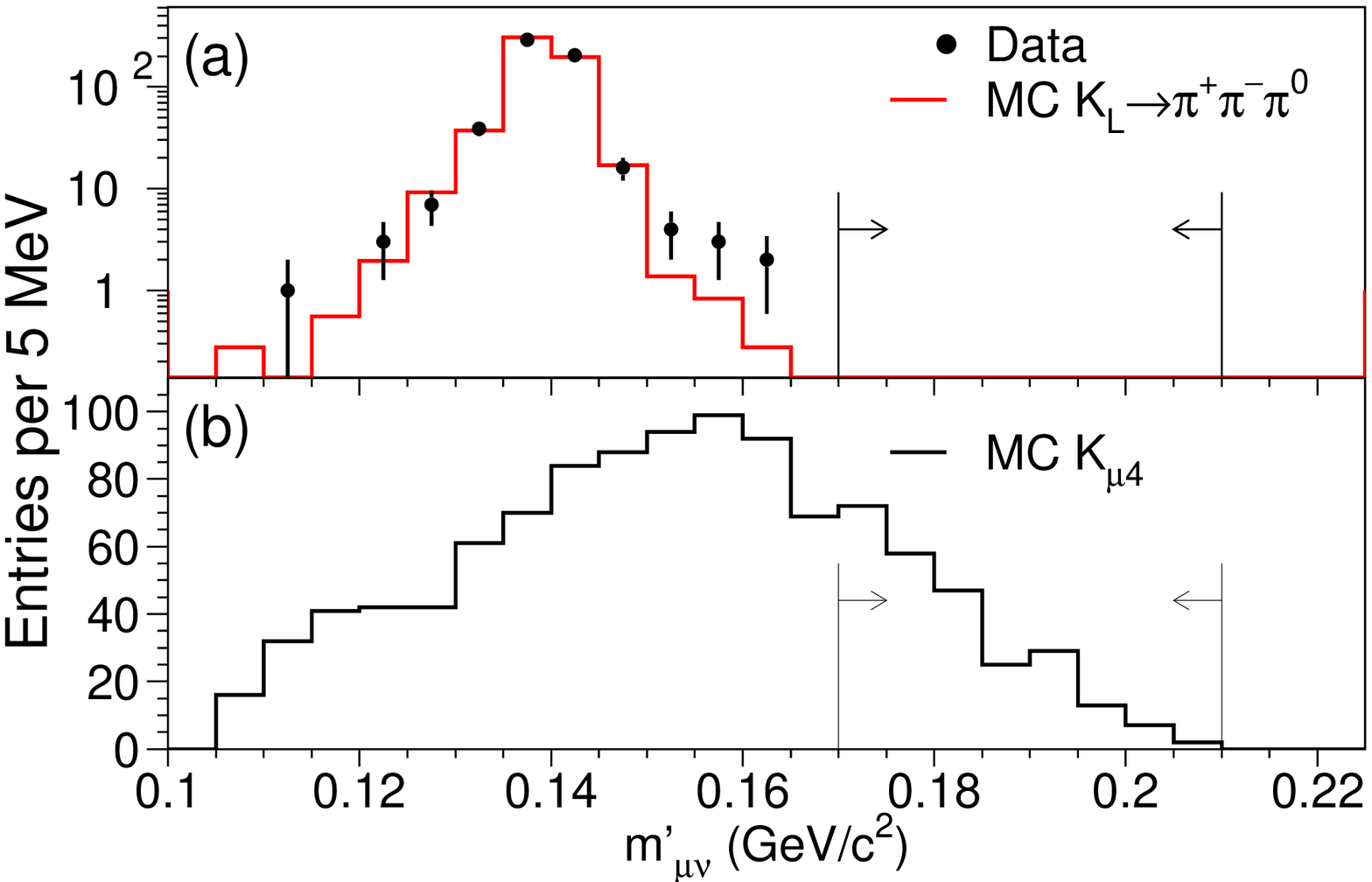,width=\linewidth}
  \caption{
    For the two kaon energy solutions,
    the invariant muon-neutrino mass closest to pion mass ($\mmunu'$) 
    is shown in (a) for
    data (dots) and MC \KLpmz\ (histogram).
    (b) shows the $\mmunu'$ distribution for MC $\km4$.
    The vertical lines show the selection window for
    $\km4$ candidates. All $\km4$ selection requirements,
    except for $\mmunu'$, have been applied.
          }
  \label{fig:mmunu}
\end{figure}

The search for $\km4$ decays uses the same data sample
as the \rkmurad\ analysis.
A pion and muon track are searched using
the two-track analysis described in Sec.~\ref{subsec:kl3_sel}.
To increase the acceptance,
requirements on the decay vertex and kaon energy are relaxed.
Next, we search for two photons in the CsI calorimeter 
such that the two-photon invariant mass is consistent with
the $\pz$ mass.  This sample of ``$\pz\pi^{\pm}\mu^{\mp}$'' candidates
is mainly background from \KLpmz\ decays in which one of the pions
decays into a muon plus neutrino.
To reduce this background,
the muon-neutrino invariant mass ($\mmunu$) is computed for 
both kaon energy solutions,
and both $\mmunu$ solutions are required to be well away
from the charged pion mass.
A $\km4$ candidate is defined such that the $\mmunu$ solution
closest to the pion mass ($\mmunu'$)
is between 170 and 210~MeV/$c^2$.
Figure~\ref{fig:mmunu}a shows the $\mmunu'$  distribution;
the data are shown as dots, and \KLpmz\ MC is shown by
the histogram. 
For data, there are no events in the signal region;
the background prediction from \KLpmz\ MC 
describes the data well. 
Figure~\ref{fig:mmunu}b
shows the $\mmunu'$ distribution for $\km4$ MC.
From this MC sample, the acceptance is determined to be 0.25\%.
For the normalization mode, there are 1.86 million \KLpimunu\ 
candidates, and the acceptance (from MC) is 13.4\%.
The resulting upper limit is
\begin{equation}
   B(\km4) < 2\times 10^{-5}~~~(90\%~{\rm confidence}).
\end{equation}

 \section{Measurement of Fake Photon Probability From Pion Interactions }
 \label{app:pihad}

A potential background for \Klthreerad\  is a non-radiative
\Klthree\ decay in which the pion interacts 
hadronically in the material upstream
of the CsI and generates a CsI cluster that satisfies the photon
identification cuts.  In this appendix, we first describe how
this effect is simulated, and then use \KLSpp\ events to measure
the fake-photon probability in both the data and simulation.

The effects of {\pihad s} are simulated with a {\sc geant}-based library
that stores hadronic secondaries produced from
pion interactions.
When a charged pion is traced through the detector in the \ktev\ MC, 
the hadronic interaction probability is computed
for each detector element.
To simulate a  \pihad,
a shower is selected from the {\sc geant} library
based on the pion momentum, and the secondaries are
traced through the detector.

To check the simulation of hadronic showers,
we measure the fake-photon probability for $\KLSpp$ decays.
This decay mode is ideal because simple kinematic cuts (see below)
remove events with radiative photons, and therefore any extra cluster
in the CsI calorimeter must be from a hadronic interaction.
To have sufficient statistics, we use the $\ppc$ sample from the
$\reepoe$~analysis\cite{ktev03_reepoe}. 
For this data sample, a regenerator placed in one of the beams
was used to generate $K_S$; $\KLSpp$ decays from both beams are
used in this study. The regenerator veto is used to suppress
accidental photon clusters arising from interactions of the 
neutral beam in the regenerator.
Very tight requirements are made on the
\ppc\ invariant mass (495-502~MeV/$c^2$) and transverse momentum
($p_t^2 < 10^{-4}$~GeV$^2/c^2$) to ensure that photons from
radiative $\KLSpp\gamma$ are well below the 3~GeV
cluster-energy requirement.

To remove events in which both pions undergo a hadronic shower,
one of the pions is required to deposit energy in only one CsI crystal
(compared with 49 crystals used to sum energy for photon showers),
and no other cluster is allowed to lie within 30~cm of this
``non-shower cluster.'' 
The probability for a pion to satisfy this non-shower requirement
is 10\%; the event is rejected if both pions satisfy the
non-shower requirement.
Finally, to reduce accidentals, the regenerator veto 
is applied during a 150~ns window (8 RF buckets) centered on
the event start time.

\begin{figure}[hb]
  \centering
  \epsfig{file=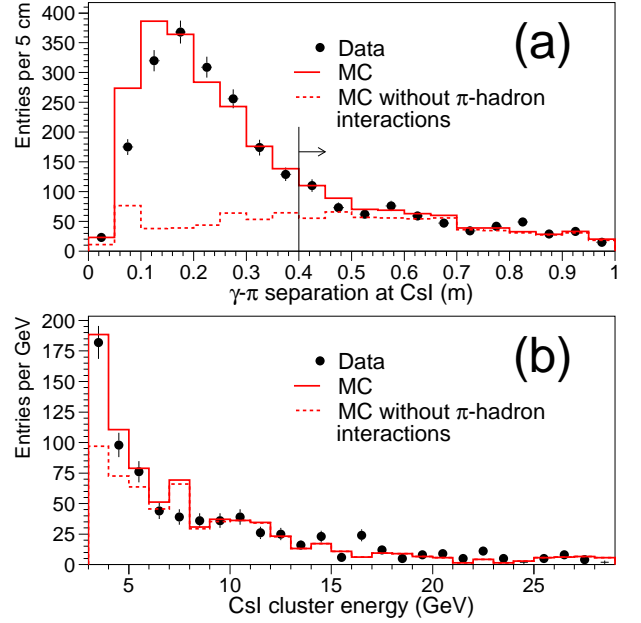,width=\linewidth}
  \caption{
    For reconstructed \KLSpp\ decays with an extra
    cluster that satisfies the \Klthreerad\ photon 
     identification requirements,
     (a) $\pi$-$\gamma$ separation at CsI 
     (with separation cut removed) and
    (b) photon energy distribution.
    Samples shown are data (dots), MC (histogram),
    and MC without \pihad s (dashed-histogram).
    The MC is normalized to the total number of \KLSpp\
    candidates.
          }
  \label{fig:2pimip}
\end{figure}

After the selection criteria described above, 
there are $4.7\times 10^6$ $\KLSpp$ candidates. 
Using this sample, photon cluster candidates
are searched for using the same criteria as in the \Klthreerad\
analysis.  
Figure~\ref{fig:2pimip}a shows the $\pi$-cluster separation
at the CsI calorimeter; the enhancement below 40~cm
is from \pihad s. 
After the 40~cm $\pi$-cluster separation requirement,
Fig.~\ref{fig:2pimip}b shows the energy distribution for
CsI clusters that satisfy the \Klthreerad\ photon requirements.
The MC prediction is also shown after normalizing the MC
to have the same number of $\KLSpp$ events as the data.
Since 98\% of the photon clusters above 6~GeV are from 
accidentals (dashed histogram in Fig~\ref{fig:2pimip}b),
this analysis uses photon cluster candidates with energy between
3 and 6~GeV.
With this additional cluster energy requirement,
the fraction of \KLSpp\ candidates 
satisfying the \Klthreerad\ photon requirements is
$\FAKERADDATA\times 10^{-5}$ in data, and
$\FAKERADMC\times 10^{-5}$ in MC.
According to the MC, 60\% of the events with extra clusters
are due to accidental activity and the remaining 40\% is from
pion interactions upstream of the CsI. 
Note that the effect of accidentals is more significant in this 
\KLSpp\ sample because the beam
intensity is much higher compared to the sample used to measure
\rklrad.
Assuming no uncertainty in the accidental
contribution for $\KLSpp$,
the relative error on the contribution from \pihad s
is 20\%.

%
%

\clearpage

\end{document}